\documentclass[12pt]{article}

\usepackage[margin=1in]{geometry}
\usepackage{setspace}
\usepackage{amsmath,amssymb,amsfonts}
\usepackage{graphicx}
\usepackage{booktabs,tabularx}
\usepackage{natbib}
\usepackage{float}
\usepackage{hyperref}
\usepackage{xcolor}
\usepackage{enumitem}
\usepackage{listings}
\usepackage{caption}
\usepackage{subcaption}
\usepackage{titling}

\newcommand{\E}{\mathbb{E}}

\newcommand{\R}{\texttt{R}}
\newcommand{\pkg}[1]{\textbf{#1}}
\newcommand{\fn}[1]{\texttt{#1}}
\newcommand{\code}[1]{\texttt{#1}}

\pretitle{\begin{center}\LARGE}
\posttitle{\par\end{center}\vspace{2.2em}}
\preauthor{\begin{center}\large \lineskip 0.5em\begin{tabular}[t]{c}}
\postauthor{\end{tabular}\par\end{center}\vspace{1.8em}}
\predate{\begin{center}\large}
\postdate{\par\end{center}\vspace{1.2em}}

\begin{document}

\title{\pkg{ForeComp}: An R Package for Comparing Predictive Accuracy Using Fixed-Smoothing Asymptotics}

\author{
Minchul Shin\thanks{\textbf{Disclaimer}: The views expressed in these papers are solely those of the authors and do not necessarily reflect the views of the Federal Reserve Bank of Philadelphia or the Federal Reserve System. Any errors or omissions are the responsibility of the authors.}\\
{\normalsize Federal Reserve Bank of Philadelphia}
\and
Nathan Schor\\
{\normalsize Baltimore Orioles}
}

\date{\today}

\maketitle

\begin{abstract}
\noindent
We introduce \pkg{ForeComp}, an \R{} package for comparing predictive accuracy using Diebold--Mariano type tests of equal predictive ability with standard and fixed-smoothing inference. The package provides a common interface for loss-differential based testing and includes \fn{Plot\_Tradeoff}, a visual diagnostic for bandwidth sensitivity and the size--power tradeoff. We illustrate the toolkit with Survey of Professional Forecasters applications and Monte Carlo evidence on finite-sample performance.



\bigskip
\noindent \textbf{Contact:} Minchul Shin (\texttt{visiblehand@gmail.com})
\end{abstract}

\bigskip

\newpage

\onehalfspacing

\section{Introduction}
\label{sec:intro}

Comparing the predictive accuracy of competing forecasts is a fundamental task in empirical economics and finance. The seminal test of \citet{DieboldMariano1995} (hereafter DM) provides a simple and widely used framework: given a sequence of loss differentials between two forecasts, the DM test assesses whether the expected loss differential is zero. The test statistic is straightforward to compute, requires minimal distributional assumptions, and is applicable to a broad range of loss functions.

Despite its simplicity, the DM test is well known to suffer from size distortions in finite samples, particularly when the evaluation sample size is small. The primary source of distortion is the estimation of the long-run variance of the loss differential. The original DM recommendation uses a rectangular kernel truncated at $h-1$ lags, where $h$ is the forecast horizon, exploiting the theoretical MA$(h-1)$ structure of optimal forecast errors. In practice, however, the loss differential may exhibit serial correlation beyond $h-1$ lags, for example when forecasts are suboptimal, when in-sample parameter estimation introduces additional dependence, or when models are misspecified. In short evaluation samples, empirically relevant bandwidth choices can change both the size of the test and the resulting rejection decision.

A growing literature has proposed alternative inference procedures that deliver more reliable size control, especially in small samples. These include the modified DM test of \citet{HarveyLeybourneNewbold1997}, the use of positive-semidefinite kernels with data-dependent bandwidth selection \citep{NeweyWest1994}, and fixed-smoothing approaches. Fixed-smoothing asymptotics, developed by \citet{KieferVogelsang2005} for the fixed-$b$ case and \citet{Sun2013} for the fixed-$m$ case, derive the limiting distribution of the test statistic under the assumption that the smoothing parameter does not vanish with sample size. The resulting reference distributions account for the estimation error in the long-run variance, improving size properties in short samples. \citet{LLSW2018} provide practical recommendations for fixed-smoothing inference and emphasize the size--power tradeoff inherent in bandwidth choice.

The \pkg{ForeComp} package for \R{} \citep{Rlang} consolidates these approaches into a single toolkit under a common interface. Table~\ref{tab:methods} summarizes the available procedures, spanning classical DM inference, fixed-$b$ and fixed-$m$ methods, orthonormal-series long-run variance estimation, and clustering-based approaches. The main testing functions take a vector of loss differentials as input and return a test statistic and rejection decision (and $p$-values where available), making it easy to compare conclusions across standard and fixed-smoothing procedures. Beyond offering multiple tests, \pkg{ForeComp} provides a visual diagnostic, \fn{Plot\_Tradeoff}, that assesses bandwidth robustness by plotting the implied size distortion--power loss tradeoff across a grid of bandwidth values and indicating where the null is rejected. This diagnostic helps users distinguish robust findings from bandwidth-driven rejections.

We illustrate the package with three empirical applications using the Survey of Professional Forecasters, including replications of \citet{Stark2010} and \citet{CoroneoIacone2020} and a bandwidth sensitivity analysis. We also report a Monte Carlo study based on \citet{McCracken2019} that summarizes finite-sample size and size-corrected power across the main procedures implemented in \pkg{ForeComp}. The simulations show that traditional normal-approximation procedures can substantially over-reject in short samples, while fixed-smoothing methods deliver improved size control with competitive size-corrected power.

The remainder of this paper is organized as follows. Section~\ref{sec:framework} establishes the testing framework. Section~\ref{sec:methods} describes the testing procedures available in \pkg{ForeComp}. Section~\ref{sec:usage} illustrates the common interface and the \fn{Plot\_Tradeoff} diagnostic. Sections~\ref{sec:app1}--\ref{sec:app3} present three empirical applications, Section~\ref{sec:mc} reports a Monte Carlo simulation study of size and power, and Section~\ref{sec:conclusion} concludes. Complete simulation tables are collected in Appendix~\ref{sec:appendix_tables}.

\section{Framework: Comparing Two Forecasts}
\label{sec:framework}

Consider a scalar target variable $y_{t+h}$ for which two competing $h$-step-ahead forecasts, $\hat{y}^{(1)}_{t+h|t}$ and $\hat{y}^{(2)}_{t+h|t}$, are available. Let $L(\cdot)$ denote a loss function (e.g., squared error), and define the \emph{loss differential}
\begin{equation}
  d_{t+h} \;=\; L\!\left(y_{t+h},\, \hat{y}^{(1)}_{t+h|t}\right) - L\!\left(y_{t+h},\, \hat{y}^{(2)}_{t+h|t}\right), \quad t = 1, 2, \ldots, P,
\label{eq:loss_diff}
\end{equation}
where $P$ denotes the number of out-of-sample evaluation observations. A positive $d_{t+h}$ indicates that forecast~1 incurs a larger loss than forecast~2 at time $t+h$.

The null hypothesis of \emph{equal predictive ability} (EPA) is
\begin{equation}
  H_0 \colon \E[d_{t+h}] = 0 \qquad \text{versus} \qquad H_1 \colon \E[d_{t+h}] \neq 0.
\label{eq:null}
\end{equation}

Under mild regularity conditions---stationarity and ergodicity of $\{d_{t+h}\}$, and the existence of a positive long-run variance $\sigma^2 = \sum_{j=-\infty}^{\infty} \gamma_j > 0$ where $\gamma_j = \mathrm{Cov}(d_{t+h},\, d_{t+h+j})$---the DM test statistic takes the form
\begin{equation}
  \mathrm{DM} = \sqrt{P}\,\frac{\bar{d}}{\hat{\sigma}} \;\xrightarrow{d}\; N(0,1) \quad \text{as } P \to \infty,
\label{eq:dm_stat}
\end{equation}
where $\bar{d} = P^{-1}\sum_{t=1}^{P} d_{t+h}$ is the sample mean of the loss differentials and $\hat{\sigma}$ is a consistent estimator of the long-run standard deviation $\sigma$. The critical challenge lies in the construction of $\hat{\sigma}$.

The generic long-run variance estimator based on weighted autocovariances is
\begin{equation}
  \hat{\sigma}^2 = \hat{\gamma}_0 + 2\sum_{j=1}^{P-1} k\!\left(\frac{j}{M}\right) \hat{\gamma}_j,
\label{eq:lrv}
\end{equation}
where $\hat{\gamma}_j = P^{-1}\sum_{t=1}^{P-j}(d_t - \bar{d})(d_{t+j} - \bar{d})$ is the sample autocovariance at lag $j$, $k(\cdot)$ is a kernel function, and $M$ is a bandwidth parameter. The choice of kernel, bandwidth, and asymptotic approximation jointly determine the finite-sample properties of the resulting test.

\section{Testing Procedures in \pkg{ForeComp}}
\label{sec:methods}

Table~\ref{tab:methods} provides a summary of all testing procedures implemented in \pkg{ForeComp}. We organize the discussion into two groups: standard procedures that use normal or $t$-distribution critical values with small bandwidths (Section~\ref{sec:standard}), and alternatives based on fixed-smoothing asymptotics or nonparametric methods (Section~\ref{sec:fixedsmoothing}).

\begin{table}[t]
\centering
\caption{Testing procedures implemented in \pkg{ForeComp}}
\label{tab:methods}
\small
\begin{tabular}{@{}llllll@{}}
\toprule
Label & Function & Kernel & Bandwidth & Critical value \\
\midrule
\multicolumn{5}{@{}l}{\textit{Standard procedures}} \\[3pt]
DM-R & \fn{dm.test.r()} & Rectangular & $M = h-1$ & $N(0,1)$ \\
DM-M & \fn{dm.test.r.m()} & Rectangular & $M = h-1$ & $t_{P-1}$ \\
DM-NW & \fn{dm.test.bt()} & Bartlett & NW(1994) & $N(0,1)$ \\
DM-NW-L & \fn{dm.test.bt(Mopt=1)} & Bartlett & LLSW(2018) & $N(0,1)$ \\
\midrule
\multicolumn{5}{@{}l}{\textit{Fixed-smoothing and alternative procedures}} \\[3pt]
DM-FB & \fn{dm.test.bt.fb()} & Bartlett & LLSW(2018) & Fixed-$b$ \\
DM-EWC & \fn{dm.test.ewc.fb()} & EWC & LLSW(2018) & $t_{B}$ \\
DM-WPE & \fn{dm.test.wpe.fb()} & Daniell & CI(2020) & $t_{2m}$ \\
DM-IM & \fn{dm.test.im()} & --- & $q$ blocks & $t_{q-1}$ \\
\bottomrule
\end{tabular}
\begin{flushleft}
\footnotesize
\textit{Notes:} NW(1994) = \citet{NeweyWest1994}, $M = \lceil 4(P/100)^{2/9}\rceil$. LLSW(2018) = \citet{LLSW2018}. CI(2020) = \citet{CoroneoIacone2020}. EWC = equal-weighted cosine estimator. The package also provides an experimental implementation of the \citet{CanayRomanoShaikh2017} randomization test (\fn{dm.test.cnr.t}, \fn{dm.test.cnr.w}); see the package documentation for details.
\end{flushleft}
\end{table}

\subsection{Standard Procedures}
\label{sec:standard}

\paragraph{DM-R: Original DM test (\fn{dm.test.r}).}
The original recommendation of \citet{DieboldMariano1995} exploits the theoretical result that, under the null hypothesis with an optimal $h$-step-ahead forecast, the loss differential follows an MA$(h-1)$ process. This motivates using the rectangular kernel with truncation at $M = h-1$:
\begin{equation}
  \hat{\sigma}^2_{\mathrm{R}} = \hat{\gamma}_0 + 2\sum_{j=1}^{h-1} \hat{\gamma}_j.
\label{eq:rect_var}
\end{equation}
The test statistic $\mathrm{DM}_R = \sqrt{P}\,\bar{d}\,/\,\hat{\sigma}_{\mathrm{R}}$ is compared to standard normal critical values. When $h=1$ (one-step-ahead forecasts), the long-run variance reduces to the sample variance. This estimator is not guaranteed to be nonnegative.

\paragraph{DM-M: Modified DM test (\fn{dm.test.r.m}).}
\citet{HarveyLeybourneNewbold1997} proposed two finite-sample corrections. First, a bias correction factor:
\begin{equation}
  \mathrm{DM}_M = \left[\frac{P + 1 - 2h + P^{-1}h(h-1)}{P}\right]^{1/2} \times \mathrm{DM}_R.
\end{equation}
Second, the test statistic is compared to $t_{P-1}$ critical values rather than standard normal, providing heavier tails that partially account for estimation uncertainty.

\paragraph{DM-NW: Bartlett kernel with normal approximation (\fn{dm.test.bt}).}
To avoid the risk of negative variance estimates inherent in the rectangular kernel, one may use the Bartlett (triangular) kernel:
\begin{equation}
  k_{\mathrm{B}}(x) = (1 - |x|)\,\boldsymbol{1}\!\left\{|x| \leq 1\right\},
\end{equation}
which guarantees $\hat{\sigma}^2 \geq 0$. The default bandwidth follows \citet{NeweyWest1994}:
\begin{equation}
  M_{\mathrm{NW}} = \left\lceil 4\left(\frac{P}{100}\right)^{2/9}\right\rceil.
\end{equation}
Standard normal critical values are used. While the positive-semidefiniteness of the Bartlett kernel is advantageous, the relatively small NW bandwidth may fail to capture serial dependence beyond $h-1$ lags, leading to size distortions similar to DM-R.

\subsection{Fixed-Smoothing and Alternative Procedures}
\label{sec:fixedsmoothing}

\paragraph{DM-FB: Bartlett kernel with fixed-$b$ asymptotics (\fn{dm.test.bt.fb}).}
The fixed-$b$ approach of \citet{KieferVogelsang2005} uses the same Bartlett kernel but with a substantially larger bandwidth,
\begin{equation}
  M_{\mathrm{FB}} = \lceil 1.3\sqrt{P}\rceil,
\label{eq:M_fb}
\end{equation}
following the recommendation of \citet{LLSW2018}. Their bandwidth rule is motivated as a practical default that approximately minimizes a loss function trading off size distortions against power loss for fixed-$b$ tests, rather than as an MSE-optimal choice for long-run variance estimation. In \pkg{ForeComp}, this LLSW rule is the default for \fn{dm.test.bt.fb} when $M$ is not supplied. The key difference from DM-NW is the asymptotic theory: letting $b = M/P$ remain fixed as $P \to \infty$, the test statistic converges to a nonstandard distribution $\Phi_{\mathrm{BART}}(b)$ whose quantiles depend on $b$. For the Bartlett kernel, \citet{KieferVogelsang2005} provide the approximation:
\begin{equation}
  q_{0.975}(b) = 1.9600 + 2.9694\,b + 0.4160\,b^2 - 0.5324\,b^3,
\label{eq:fixedb_cv}
\end{equation}
for the 97.5\% quantile (used in a two-sided 5\% test). The larger bandwidth captures more serial dependence, while the adjusted critical values account for the inconsistency of the variance estimator, jointly producing much better size control.

\paragraph{DM-EWC: Equal-weighted cosine estimator (\fn{dm.test.ewc.fb}).}
The equal-weighted cosine (EWC) estimator of \citet{LLSW2018} is an orthonormal series estimator of the long-run variance:
\begin{equation}
  \hat{\lambda}_j = \sqrt{\frac{2}{P}}\sum_{t=1}^{P} d_t \cos\!\left[\pi j\!\left(\frac{t - 1/2}{P}\right)\right], \qquad
  \hat{\sigma}^2_{\mathrm{EWC}} = \frac{1}{B}\sum_{j=1}^{B} \hat{\lambda}_j^2,
\end{equation}
with bandwidth $B = \lfloor 0.4\,P^{2/3}\rfloor$. In \citet{LLSW2018}, this default is motivated by the same size--power tradeoff criterion used to select the Bartlett fixed-$b$ bandwidth. Under fixed-$b$ asymptotics, the test statistic follows a $t_B$ distribution \citep{Muller2004}, making critical values readily available from standard tables.

\paragraph{DM-WPE: Weighted periodogram estimator with fixed-$m$ asymptotics (\fn{dm.test.wpe.fb}).}
\citet{CoroneoIacone2020} apply the weighted periodogram estimator (WPE) with the Daniell kernel to forecast comparison:
\begin{equation}
  \hat{\sigma}^2_{\mathrm{WPE}} = \frac{2\pi}{m}\sum_{j=1}^{m} I(\lambda_j), \qquad \lambda_j = \frac{2\pi j}{P},
\label{eq:wpe}
\end{equation}
where $I(\lambda_j) = |{(2\pi P)^{-1/2}\sum_{t=1}^{P} d_t e^{-i\lambda_j t}}|^2$ is the periodogram and $m = \lfloor P^{1/3}\rfloor$ is the bandwidth. Under fixed-$m$ asymptotics \citep{Sun2013}, the test statistic $\sqrt{P}\,\bar{d}\,/\,\hat{\sigma}_{\mathrm{WPE}}$ follows a $t_{2m}$ distribution.

\paragraph{DM-IM: Ibragimov--M\"uller test (\fn{dm.test.im}).}
\citet{IbragimovMuller2010} propose a clustering approach that avoids explicit long-run variance estimation. The evaluation sample is divided into $q$ nonoverlapping blocks. The block means $\bar{d}_1, \ldots, \bar{d}_q$ are treated as $q$ independent observations, and a standard $t$-test is applied:
\begin{equation}
  \mathrm{DM}_{\mathrm{IM}} = \frac{\bar{m}}{\sqrt{s^2_q / q}}, \qquad s^2_q = \frac{1}{q-1}\sum_{j=1}^{q}(\bar{d}_j - \bar{m})^2,
\end{equation}
where $\bar{m} = q^{-1}\sum_{j=1}^{q}\bar{d}_j$. The critical values come from the $t_{q-1}$ distribution. The approach is simple and robust but may lack power due to the small effective sample size $q$.

\section{Package Usage}
\label{sec:usage}

\subsection{Installation}

The \pkg{ForeComp} package can be installed from the package source:
\begin{lstlisting}
# Install from local source
install.packages("ForeComp", repos = NULL, type = "source")

# Load the package
library(ForeComp)
\end{lstlisting}

\subsection{Common Interface}

All testing functions in \pkg{ForeComp} accept a numeric vector \code{d} of loss differentials as their first argument and return a list with elements \code{\$rej} (logical rejection decision), \code{\$stat} (test statistic), and \code{\$pval} ($p$-value, where available). The significance level is controlled by the argument \code{cl} (default 0.05).

A typical workflow proceeds in two steps. As a running example, we compare the SPF consensus nowcast ($h = 0$) of real output growth against the no-change benchmark over 2007:Q1--2016:Q4 ($T = 40$). First, compute the loss differential from the forecast errors:
\begin{lstlisting}
library(ForeComp)
df = RGDP[RGDP$X1 >= "2007:01" & RGDP$X1 <= "2016:04", ]

# Forecast errors
e_nc  = df$Realiz1 - df$NCfor_Step1
e_spf = df$Realiz1 - df$SPFfor_Step1

# Squared-error loss differential
d = e_nc^2 - e_spf^2
d = d[!is.na(d)]  # T = 40
\end{lstlisting}

Second, apply the desired test:
\begin{lstlisting}
result = dm.test.r(d, h = 1, cl = 0.05)
result$stat   # 2.79
result$pval   # p-value
result$rej    # TRUE (reject H0 at 5%)
\end{lstlisting}

\subsection{Function Reference}
\label{sec:functions}

Table~\ref{tab:functions} summarizes the key arguments for each function.

\begin{table}[t]
\centering
\caption{Function arguments in \pkg{ForeComp}}
\label{tab:functions}
\small
\begin{tabular}{@{}lp{9cm}@{}}
\toprule
Function & Key arguments \\
\midrule
\fn{dm.test.r(d, h, cl)} & \code{h}: forecast horizon (default 1). Truncation $= h-1$. \\[4pt]
\fn{dm.test.r.m(d, h, cl)} & Same as \fn{dm.test.r} but with HLN correction and $t_{P-1}$ critical values. \\[4pt]
\fn{dm.test.bt(d, M, Mopt, cl)} & \code{M}: bandwidth (if \code{NA}, automatic). \code{Mopt}: 1 = LLSW ($M=\lceil 1.3\sqrt{P}\rceil$), 2 = NW(1994, default: $M=\lceil 4(P/100)^{2/9}\rceil$), 3 = textbook NW/Andrews ($M=\lceil 0.75P^{1/3}\rceil$), 4 = CI baseline ($M=\lfloor P^{1/2}\rfloor$). \\[4pt]
\fn{dm.test.bt.fb(d, M, Mopt, cl)} & Same kernel as \fn{dm.test.bt} but with fixed-$b$ critical values. If \code{M = NA}, default \code{Mopt = 1} (LLSW) uses $M=\lceil 1.3\sqrt{P}\rceil$; \code{Mopt = 2,3,4} use NW(1994), textbook NW, and CI baseline, respectively. \\[4pt]
\fn{dm.test.ewc.fb(d, B, Bopt, cl)} & \code{B}: bandwidth (default: $\lfloor 0.4 P^{2/3}\rfloor$). \\[4pt]
\fn{dm.test.wpe.fb(d, M, Mopt, cl)} & \code{M}: bandwidth (default: $\lfloor P^{1/3}\rfloor$). Uses $t_{2M}$ critical values. \\[4pt]
\fn{dm.test.im(d, q, cl)} & \code{q}: number of blocks (default 2). \\
\bottomrule
\end{tabular}
\end{table}

\subsection{Comparing Multiple Tests}

The common interface makes it straightforward to compare results across methods. Continuing the running example:
\begin{lstlisting}
dm.test.r(d, h = 1)$stat      # DM-R:   2.79  (reject)
dm.test.r.m(d, h = 1)$stat    # DM-M:   2.76  (reject)
dm.test.bt(d)$stat             # DM-NW:  2.24  (reject)
dm.test.bt.fb(d)$stat          # DM-FB:  2.26  (not reject)
dm.test.ewc.fb(d)$stat         # DM-EWC: 1.83  (not reject)
dm.test.wpe.fb(d)$stat         # DM-WPE: 1.98  (not reject)
dm.test.im(d, q = 5)$stat      # DM-IM:  2.15  (not reject)
\end{lstlisting}

The standard tests (DM-R, DM-M, DM-NW) all reject the null at 5\%, while every fixed-smoothing alternative (DM-FB, DM-EWC, DM-WPE, DM-IM) does not. This divergence suggests that the rejections by the standard tests may be driven by size distortion rather than by strong evidence against equal predictive ability---a pattern consistent with the Monte Carlo findings in Section~\ref{sec:mc}, which document substantial over-rejection by DM-R and DM-NW in small samples. When such a divergence arises, we recommend favoring the fixed-smoothing results, which are designed to maintain reliable size control in finite samples.

\subsection{Bandwidth Sensitivity Diagnostic}

The WCE-B test (fixed-$b$ Bartlett) requires the practitioner to choose a bandwidth parameter $M$. In \citet{CoroneoIacone2020}, a common baseline is $M = \lfloor T^{1/2} \rfloor$ (``CI baseline''); the authors justify this choice by noting that it performed well in a preliminary Monte Carlo study (see their Supporting Information Appendix). In \pkg{ForeComp}, if the user does not supply $M$, \fn{dm.test.bt.fb} defaults to the LLSW rule $M = \lceil 1.3\sqrt{T}\rceil$ (``package default''). This default is motivated by \citet{LLSW2018}'s loss-minimization criterion that trades off size distortions against power loss under fixed-$b$ inference. In practice, the rejection decision can be sensitive to the bandwidth choice: a test that rejects at one bandwidth may fail to reject at a slightly different $M$, or vice versa.

The \fn{Plot\_Tradeoff} function provides a visual diagnostic for this sensitivity. Given two sets of forecasts and a realized outcome series, it computes the WCE-B test statistic at each bandwidth $M$ in a user-specified grid (or a default grid centered around the package default), and for each $M$ it estimates the \emph{size distortion} and \emph{maximum power loss} via simulation. The size distortion at bandwidth $M$ is defined as the difference between the empirical rejection rate under the null and the nominal level (5\%), where the null distribution is approximated by fitting an ARIMA model to the demeaned loss differential and simulating from it. The maximum power loss is the largest gap between the oracle power envelope (the power of a test with known long-run variance) and the size-corrected power of the WCE-B test, maximized over a grid of local alternatives. The green marker highlights the package default bandwidth. The function returns a list containing a \code{ggplot2} plot object and the underlying data.

As an illustration, consider testing whether the SPF nowcast of real output growth outperforms the no-change benchmark in the subsample 2007:Q1--2016:Q4 ($T = 40$):
\begin{lstlisting}
library(ForeComp)
# Subset SPF data: h=0, 2007:Q1-2016:Q4
df = RGDP[RGDP$X1 >= "2007:01" & RGDP$X1 <= "2016:04",
          c("SPFfor_Step1", "NCfor_Step1", "Realiz1")]
df = df[complete.cases(df), ]

output = Plot_Tradeoff(data = df, n_sim = 5000)
output[[1]]  # size-power tradeoff plot (Figure 1)
output[[2]]  # data frame: M, size distortion, power loss
\end{lstlisting}

\begin{figure}[t!]
\centering
\includegraphics[width=0.65\textwidth]{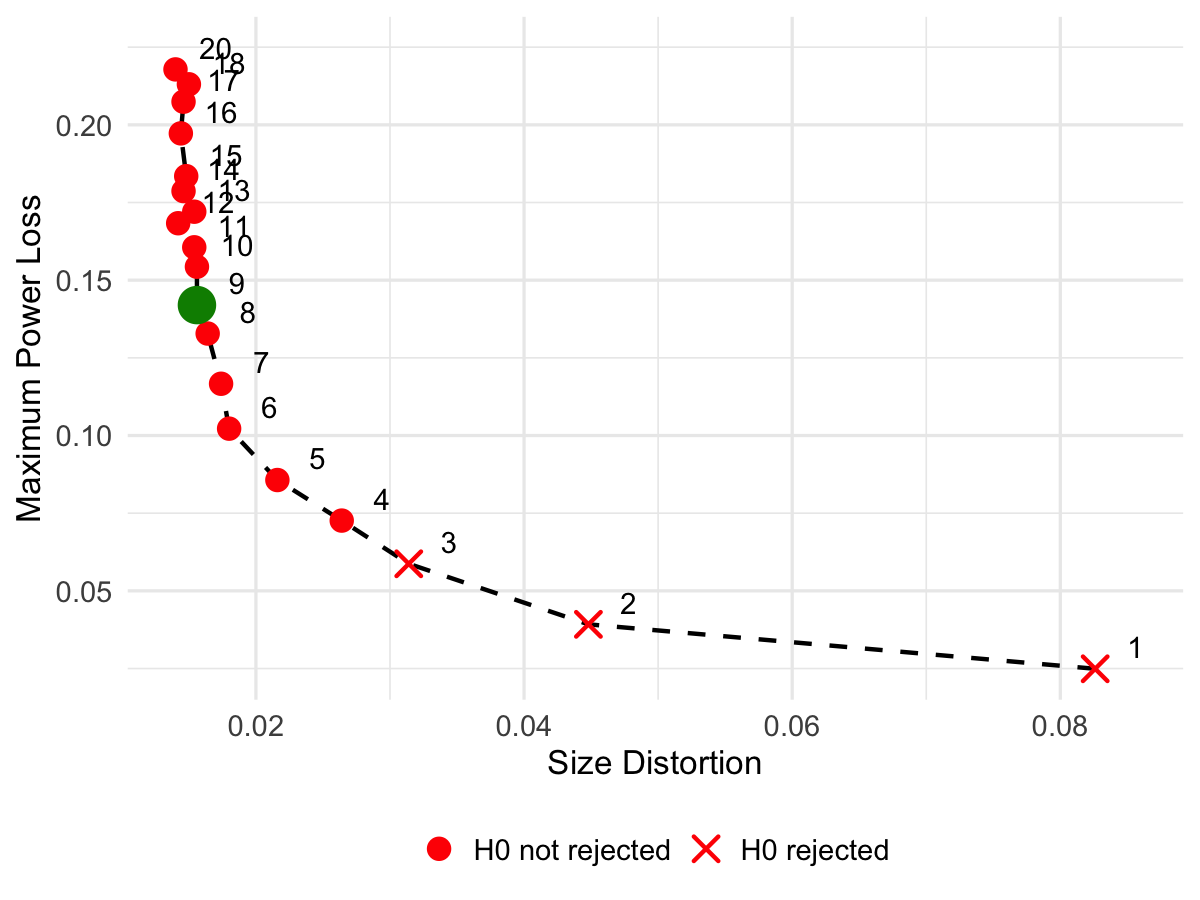}
\caption{Size--power tradeoff: Real output growth, $h = 0$, 2007:Q1--2016:Q4 ($T = 40$). Each point corresponds to a bandwidth $M$. Red circles ($\bullet$): H0 not rejected by WCE-B. Red crosses ($\times$): H0 rejected. Green marker: package default $M = \lceil 1.3\sqrt{T} \rceil = 9$.}
\label{fig:tradeoff_rgdp}
\end{figure}

Figure~\ref{fig:tradeoff_rgdp} shows the resulting tradeoff plot. Each marker on the curve corresponds to a specific bandwidth value $M$, annotated next to the point. Red crosses ($\times$) indicate bandwidths at which the WCE-B test rejects $H_0$ at the 5\% level; red circles ($\bullet$) indicate bandwidths at which it does not reject. The green marker highlights the package default $M = \lceil 1.3\sqrt{T} \rceil = 9$. As $M$ increases, the size distortion generally decreases (the test becomes more conservative) while the maximum power loss increases (the test loses ability to detect alternatives)---producing the characteristic tradeoff curve. In this example, only the smallest bandwidths ($M = 1, 2, 3$) reject the null; at the package default $M = 9$ and all larger bandwidths, the test does not reject. The non-rejection is robust across most of the frontier, and the figure makes it transparent how the practitioner could deviate from the default, moving to smaller $M$ to favor power or to larger $M$ to favor size control, while seeing exactly where the rejection decision changes. Application~3 presents additional examples using SPF data.

\section{Application 1: SPF Forecast Evaluation}
\label{sec:app1}

Our first application replicates selected results from \citet{Stark2010}, who evaluates the predictive accuracy of the Survey of Professional Forecasters (SPF) for U.S.\ real output growth. The SPF, maintained by the Federal Reserve Bank of Philadelphia, provides median forecasts at horizons from the current quarter ($h=0$) through four quarters ahead ($h=4$).

\citet{Stark2010} compares SPF median forecasts against three benchmark models---no-change (NC), indirect autoregression (IAR), and direct autoregression (DAR)---using RMSE and Diebold--Mariano tests with a Bartlett kernel. The evaluation period is 1985:Q1--2007:Q4. A distinctive feature of this analysis is the use of multiple realization vintages (initial release through latest available) to assess the sensitivity of forecast evaluation to data revisions.

Using the \pkg{ForeComp} package data and the \fn{dm.test.bt} function, we replicate Table~1 of \citet{Stark2010}. The data are loaded from the package:
\begin{lstlisting}
library(ForeComp)
df = RGDP[RGDP$X1 >= "1985:01" & RGDP$X1 <= "2007:04", ]
\end{lstlisting}

For each realization vintage (Realiz1 through Realiz5) and each forecast horizon, we compute the squared forecast errors and apply the DM test with a Bartlett kernel and truncation $M = h + 4$, following the paper's specification:
\begin{lstlisting}
e_spf = as.numeric(realiz) - as.numeric(spf)
e_nc  = as.numeric(realiz) - as.numeric(nc)
d     = e_spf^2 - e_nc^2
d     = d[!is.na(d)]  # listwise deletion of #N/A rows
test  = dm.test.bt(d, M = horizon + 4)
\end{lstlisting}

Table~\ref{tab:stark_table1} reproduces Panels~1 and~2 of \citet{Stark2010} Table~1. Panel~1 uses the initial-release realization; Panel~2 uses the realization available one quarter after initial release. We also replicate the remaining panels (not shown): Panels~1--4 match the original to the third decimal place, while Panel~5 (latest vintage) differs slightly because our data incorporate revisions beyond the 2010:Q2 vintage used in the original paper.

\begin{table}[t]
\centering
\caption{SPF RMSE statistics for output growth, 1985:Q1--2007:Q4 (replication of Stark Table~1)}
\label{tab:stark_table1}
\small
\begin{tabular}{@{}ccccc@{}}
\toprule
& & \multicolumn{3}{c}{RMSE ratio (DM $p$-value)} \\
\cmidrule(l){3-5}
Horizon $W$ & SPF RMSE & SPF/NC & SPF/IAR & SPF/DAR \\
\midrule
\multicolumn{5}{@{}l}{\textit{Panel 1: Realizations on initial release ($n = 90$)}} \\[3pt]
0 & 1.40 & 0.681 (0.000) & 0.806 (0.020) & 0.806 (0.020) \\
1 & 1.65 & 0.771 (0.001) & 0.925 (0.443) & 0.933 (0.483) \\
2 & 1.77 & 0.782 (0.016) & 0.986 (0.863) & 0.985 (0.861) \\
3 & 1.80 & 0.770 (0.025) & 0.994 (0.911) & 0.993 (0.890) \\
4 & 1.76 & 0.736 (0.004) & 0.981 (0.711) & 0.948 (0.287) \\[6pt]
\multicolumn{5}{@{}l}{\textit{Panel 2: Realizations one quarter after initial release ($n = 91$)}} \\[3pt]
0 & 1.69 & 0.778 (0.000) & 0.878 (0.070) & 0.878 (0.070) \\
1 & 1.82 & 0.826 (0.006) & 0.925 (0.370) & 0.936 (0.421) \\
2 & 1.91 & 0.785 (0.013) & 0.956 (0.566) & 0.958 (0.581) \\
3 & 1.95 & 0.758 (0.020) & 0.969 (0.582) & 0.963 (0.496) \\
4 & 1.95 & 0.742 (0.002) & 0.975 (0.612) & 0.953 (0.300) \\
\bottomrule
\end{tabular}
\begin{flushleft}
\footnotesize
\textit{Notes:} Ratios below 1 indicate the SPF outperforms the benchmark. $P$-values (in parentheses) are from two-sided DM tests using the Bartlett kernel with truncation $M = W + 4$, computed via \fn{dm.test.bt()}. NC = no-change, IAR = indirect autoregression, DAR = direct autoregression. Rows where any column contains \texttt{\#N/A} are dropped (listwise deletion).
\end{flushleft}
\end{table}

The results show that the SPF consistently outperforms the no-change benchmark (ratios well below 1 with small $p$-values), but performs similarly to the autoregressive benchmarks at longer horizons (ratios near 1 with large $p$-values). The RMSE levels increase across panels as the realization vintage moves further from initial release, reflecting the accumulation of data revisions. This application demonstrates both the direct usability of the package functions and the importance of data vintage choices in real-time forecast evaluation.

\section{Application 2: Fixed-Smoothing Asymptotics for SPF Evaluation}
\label{sec:app2}

Our second application replicates \citet{CoroneoIacone2020}, who demonstrate that standard DM tests can yield spuriously significant results in small evaluation samples, and that fixed-smoothing asymptotics provides a corrective.

\citet{CoroneoIacone2020} evaluate the predictive accuracy of the SPF and the ECB Survey of Professional Forecasters against a random walk benchmark. They apply three variants of the DM test:
\begin{itemize}[nosep]
  \item \textbf{WCE-DM}: Rectangular kernel with truncation equal to the forecast horizon, standard normal critical values (equivalent to our DM-R with $h = \text{horizon} + 1$).
  \item \textbf{WCE-B}: Bartlett kernel with $M = \lfloor T^{1/2}\rfloor$ (CI baseline) and fixed-$b$ critical values (our \fn{dm.test.bt.fb} with explicit $M$).
  \item \textbf{WPE-D}: Daniell kernel with $m = \lfloor T^{1/3}\rfloor$ and fixed-$m$ asymptotics (our \fn{dm.test.wpe.fb} with explicit $M$).
\end{itemize}

The evaluation covers four U.S.\ macroeconomic variables (real output growth, inflation, unemployment, and the 3-month Treasury bill rate) over 1987:Q1--2016:Q4 ($T = 120$) and three 10-year subsamples ($T = 40$ each), plus ECB SPF forecasts of GDP growth and HICP inflation over 2006:Q1--2016:Q4 ($T = 44$).

A key methodological choice is the handling of missing values. Observations where the forecast error is undefined (due to \code{\#N/A} entries in the SPF data) are set to zero rather than dropped, preserving the sample sizes reported in the paper. Additionally, only the first-release realization (Realiz1) is used for all horizons---the Realiz1--Realiz5 columns represent different data \emph{vintages}, not different forecast horizons.

\begin{lstlisting}
# Forecast errors (NAs set to 0, matching CI's approach)
e_nc  = as.numeric(realiz1) - as.numeric(nc)
e_spf = as.numeric(realiz1) - as.numeric(spf)
e_nc[is.na(e_nc)]   = 0
e_spf[is.na(e_spf)] = 0
d = e_nc^2 - e_spf^2

# Three test variants
n = length(d)
dm_r   = dm.test.r(d, h = step)
dm_fb  = dm.test.bt.fb(d, M = floor(sqrt(n)))
dm_wpe = dm.test.wpe.fb(d, M = floor(n^(1/3)))
\end{lstlisting}

Table~\ref{tab:ci_table1} reproduces Table~1 of \citet{CoroneoIacone2020} for real output growth. The remaining four tables---GNP/GDP inflation, unemployment, the 3-month Treasury bill, and the ECB SPF---are reported in Appendix~\ref{sec:appendix_ci} (Tables~\ref{tab:ci_table2}--\ref{tab:ci_table5}). Our replication matches all 258 cells across all five tables exactly at two decimal places, confirming that the \pkg{ForeComp} implementations are consistent with the original analysis.

\begin{table}[t]
\centering
\caption{Real output growth: SPF versus random walk (replication of CI Table~1)}
\label{tab:ci_table1}
\small
\begin{tabular}{@{}lccccc@{}}
\toprule
& \multicolumn{5}{c}{Forecast horizon} \\
\cmidrule(l){2-6}
& $h=0$ & $h=1$ & $h=2$ & $h=3$ & $h=4$ \\
\midrule
\multicolumn{6}{@{}l}{\textit{Evaluation period: 1987:Q1--2016:Q4, $T = 120$}} \\[3pt]
WCE-DM             & 4.85 & 2.78 & 1.84 & 1.46 & 2.13 \\
WCE-B, $M = \lfloor T^{1/2}\rfloor$  & 4.47** & 2.64** & 1.96* & 1.76 & 2.24** \\
WPE-D, $m = \lfloor T^{1/3}\rfloor$  & 5.21** & 2.63** & 1.86* & 1.68 & 2.08* \\[6pt]
\multicolumn{6}{@{}l}{\textit{Evaluation period: 1987:Q1--1996:Q4, $T = 40$}} \\[3pt]
WCE-DM             & 3.61 & 2.28 & 2.51 & 1.51 & 3.35 \\
WCE-B              & 3.92** & 2.18* & 2.52** & 1.68 & 2.01* \\
WPE-D              & 4.01** & 2.01* & 2.72** & 1.55 & 2.38* \\[6pt]
\multicolumn{6}{@{}l}{\textit{Evaluation period: 1997:Q1--2006:Q4, $T = 40$}} \\[3pt]
WCE-DM             & 2.27 & 3.21 & 1.37 & 0.87 & 1.84 \\
WCE-B              & 2.28* & 3.01** & 1.23 & 0.94 & 1.87 \\
WPE-D              & 2.04* & 2.71** & 1.13 & 0.84 & 1.68 \\[6pt]
\multicolumn{6}{@{}l}{\textit{Evaluation period: 2007:Q1--2016:Q4, $T = 40$}} \\[3pt]
WCE-DM             & 2.79 & 1.63 & 1.14 & 0.99 & 1.12 \\
WCE-B              & 2.19* & 1.59 & 1.22 & 1.09 & 1.10 \\
WPE-D              & 1.98* & 1.43 & 1.08 & 0.94 & 0.98 \\
\bottomrule
\end{tabular}
\begin{flushleft}
\footnotesize
\textit{Notes:} WCE-DM uses a normal approximation, with two-sided 5\% and 10\% critical values 1.96 and 1.645. WCE-B (fixed-$b$) and WPE-D (fixed-$m$) use fixed-smoothing asymptotics and associated reference distributions rather than the standard normal approximation; see Section~\ref{sec:fixedsmoothing} for details. Asterisks are reported only for WCE-B and WPE-D, where ** and * denote two-sided significance at the 5\% and 10\% levels.
\end{flushleft}
\end{table}

The results illustrate the practical importance of fixed-smoothing asymptotics. For real output growth, the standard normal-approximation DM statistic (WCE-DM) and the fixed-smoothing procedures (WCE-B and WPE-D) typically deliver the same qualitative inference, especially at conventional significance levels. There are, however, a few instances where WCE-DM rejects but the fixed-smoothing procedures do not. For example, in the 1997:Q1--2006:Q4 subsample at horizon $h = 4$, WCE-DM equals 1.84 and would reject at the 10\% level under the normal approximation, while both WCE-B and WPE-D fail to reject at 10\%. Such disagreements are consistent with the tendency of normal-approximation DM tests to over-reject in short samples, as documented in our Monte Carlo results in Section~\ref{sec:mc}.

\section{Application 3: Bandwidth Sensitivity of the WCE-B Test}
\label{sec:app3}

In Application~2, we follow \citet{CoroneoIacone2020} and report WCE-B results under the CI baseline bandwidth $M = \lfloor T^{1/2} \rfloor$. In \pkg{ForeComp}, however, the default bandwidth for \fn{dm.test.bt.fb} is the LLSW rule $M = \lceil 1.3\sqrt{T}\rceil$. Since the rejection decision may depend on this choice in small samples, the \fn{Plot\_Tradeoff} function allows practitioners to visualize how the test conclusion varies across a range of bandwidths. We apply this diagnostic to three cases drawn from the SPF evaluation in Application~2, comparing the SPF consensus forecast against a na\"ive (no-change) benchmark under squared-error loss.

\paragraph{Real output growth, $h = 0$, 2007:Q1--2016:Q4 ($T = 40$).}
Figure~\ref{fig:tradeoff_rgdp} displays the size--power tradeoff for the WCE-B test applied to the SPF nowcast of real output growth in the final subsample. The test rejects the null only at the smallest bandwidths ($M = 1, 2, 3$, marked with crosses); for all remaining bandwidths, including the package default $M = 9$ (green marker), the test does not reject (circles). The non-rejection at the default bandwidth is robust: the decision does not change as $M$ increases beyond 9.

The default $M = 9$ sits on the tradeoff frontier. Moving to a slightly smaller $M$ would reduce the maximum power loss but at the cost of increased size distortion; importantly, this shift does not change the rejection decision. Conversely, moving to a slightly larger $M$ would reduce size distortion further, at the cost of additional power loss, again without changing the conclusion.

This figure illustrates a key benefit of the \fn{Plot\_Tradeoff} diagnostic: it allows the practitioner to navigate the tradeoff and deviate from the default bandwidth in a principled way. Depending on the practitioner's preference---whether to prioritize size control (by moving to larger $M$) or to preserve power (by moving to smaller $M$)---the figure makes the tradeoff explicit and enables an informed choice. In this example, the non-rejection is stable across most of the frontier, giving the practitioner confidence in the conclusion regardless of bandwidth preference.

\begin{figure}[H]
\centering
\begin{subfigure}[t]{0.48\textwidth}
  \centering
  \includegraphics[width=\textwidth]{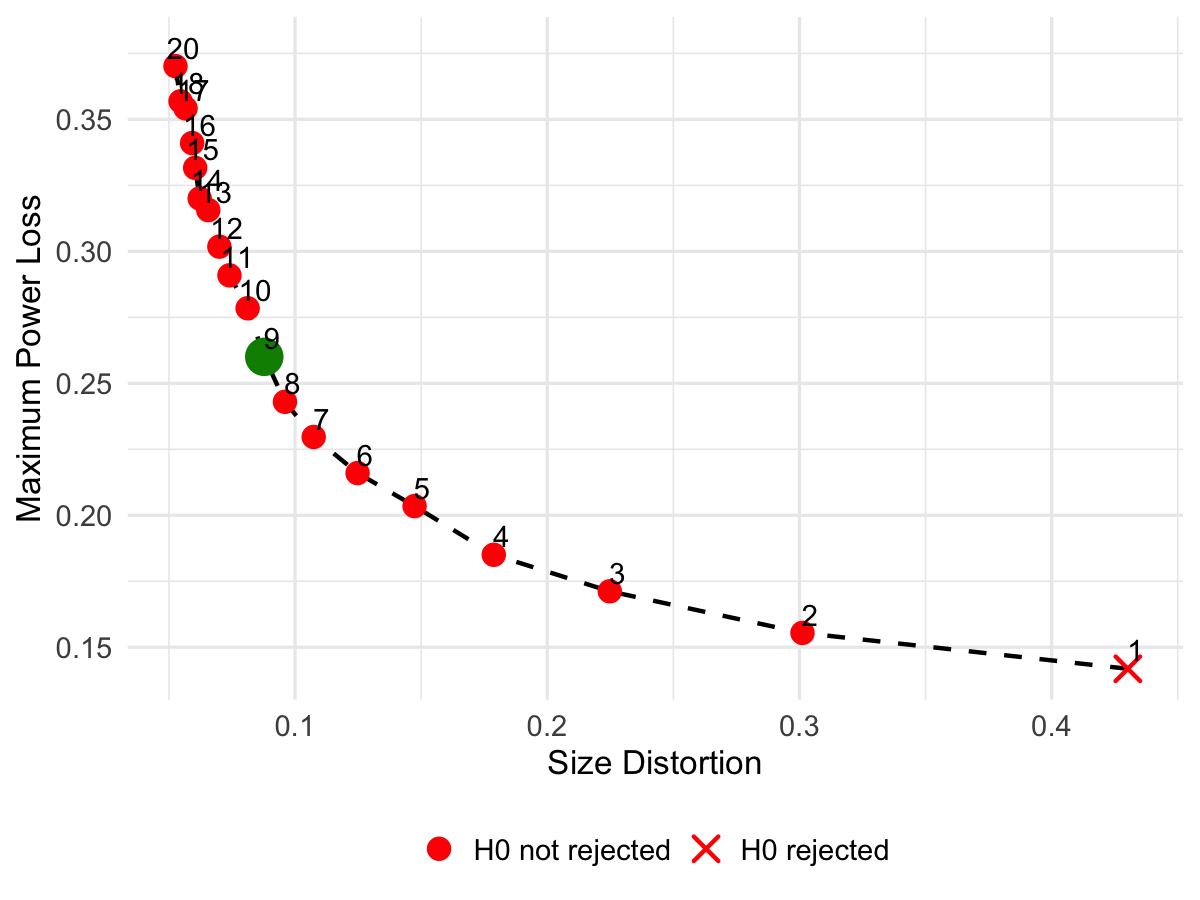}
  \caption{1997:Q1--2006:Q4}
  \label{fig:tradeoff_unemp97}
\end{subfigure}
\hfill
\begin{subfigure}[t]{0.48\textwidth}
  \centering
  \includegraphics[width=\textwidth]{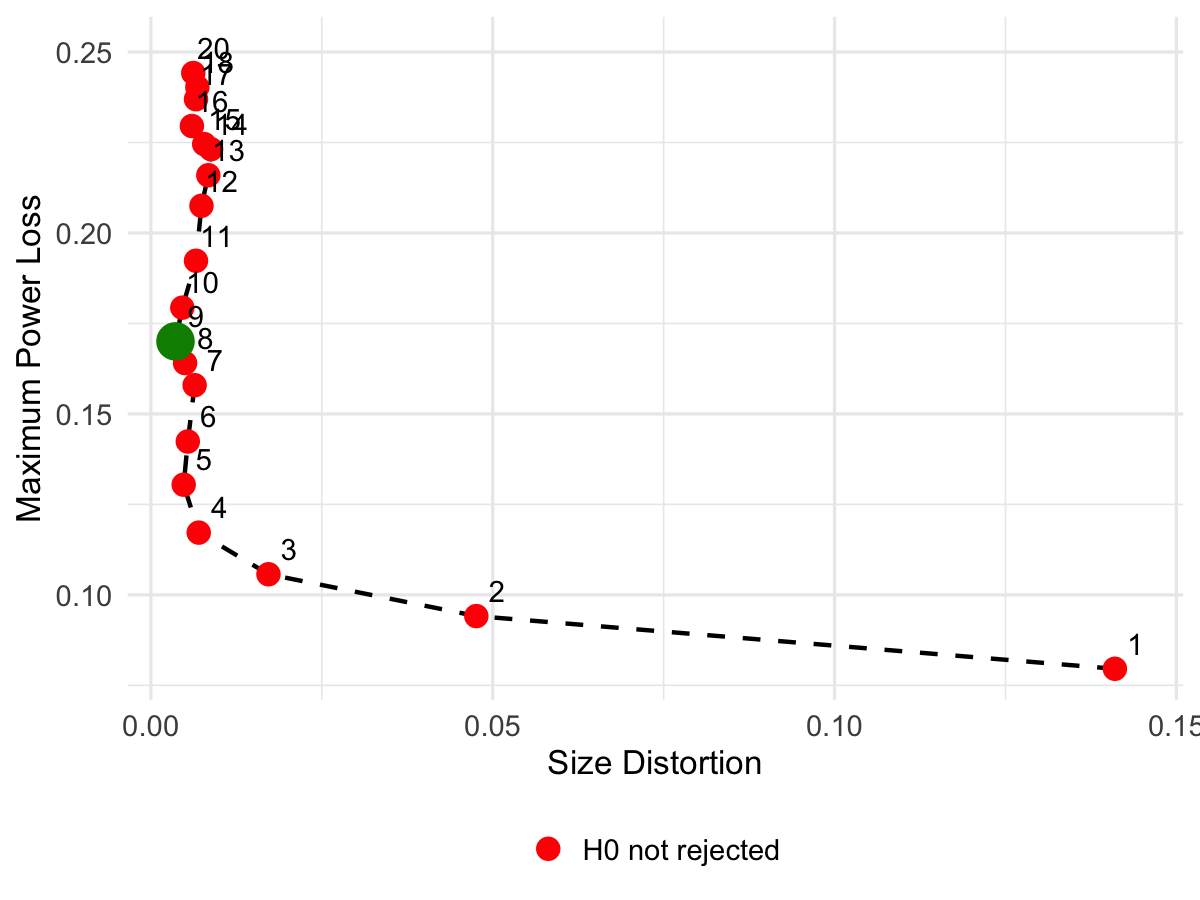}
  \caption{1987:Q1--1996:Q4}
  \label{fig:tradeoff_unemp87}
\end{subfigure}
\caption{Size--power tradeoff: Unemployment, $h = 4$, $T = 40$. Each point corresponds to a bandwidth $M$. Red circles ($\bullet$): H0 not rejected by WCE-B. Red crosses ($\times$): H0 rejected. Green marker: package default $M = \lceil 1.3\sqrt{T} \rceil = 9$.}
\label{fig:tradeoff_unemp}
\end{figure}

\paragraph{Unemployment, $h = 4$: robust non-rejection and robust rejection ($T = 40$).}
Figure~\ref{fig:tradeoff_unemp} displays the tradeoff plots for two subsamples of the unemployment forecast at the 4-quarter horizon. Panel~(a) illustrates robust non-rejection (except at the smallest bandwidth), while panel~(b) illustrates robust rejection across the bandwidth grid.

Panel~(a) shows the middle subsample (1997:Q1--2006:Q4). The WCE-B test fails to reject the null at virtually every bandwidth; only at $M = 1$ does the test reject. At that bandwidth, the Bartlett kernel collapses to a simple variance estimator and the resulting test has severe size distortions. At the default $M = 9$ and all larger bandwidths, every marker is a circle (H0 not rejected).

Panel~(b) shows the earlier subsample (1987:Q1--1996:Q4). Here the WCE-B test rejects at every bandwidth in the grid---all markers are crosses---indicating that the SPF forecast significantly outperforms the na\"ive benchmark in this subsample. Since the rejection is robust across bandwidths, the practitioner can be confident that the conclusion is not an artifact of bandwidth choice.

One difference worth highlighting is the \emph{local} shape of the tradeoff frontier around the default bandwidth (green marker). Relative to panel~(a), panel~(b) features a steeper ``cliff'': moving to slightly smaller $M$ reduces the maximum power loss (i.e., improves power) at negligible cost in terms of size distortion. In this example such a change would not overturn the test decision, but the visualization makes the practitioner's size--power tradeoff transparent and underscores that the most attractive bandwidth adjustment can be case-by-case.

These three examples illustrate the range of outcomes that \fn{Plot\_Tradeoff} can reveal. When the rejection decision changes across the bandwidth grid (Figure~\ref{fig:tradeoff_rgdp}), the test conclusion is fragile, and the practitioner should interpret the result with caution or consider the tradeoff between size control and power in choosing $M$. When the decision is stable across the entire frontier (Figure~\ref{fig:tradeoff_unemp}), either rejecting or not rejecting, the bandwidth choice is immaterial and the test provides a clear signal.

\section{Monte Carlo Simulation}
\label{sec:mc}

To systematically evaluate the finite-sample properties of the testing procedures in \pkg{ForeComp}, we conduct a Monte Carlo study based on the simulation design of \citet{McCracken2019}. We consider two data-generating processes (DGPs) that differ in the serial correlation structure of the loss differential, and we examine both empirical size and size-corrected power across a wide range of sample sizes and forecast horizons.

\subsection{Data-Generating Processes}
\label{sec:mc_dgp}

Both DGPs generate data under the null hypothesis of equal predictive ability, where a zero-forecast (Model~1: $\hat{y}^{(1)}_{t+h|t} = 0$) is compared against a rolling-mean forecast (Model~2: $\hat{y}^{(2)}_{t+h|t} = \widetilde{R}^{-1}\sum_{s=t-\widetilde{R}+1}^{t} y_s$) using squared-error loss. The parameters $R$ and $\widetilde{R}$ govern, respectively, the serial correlation structure and the rolling estimation window.

\paragraph{Unconditional-Rolling (UCR) DGP.}
The target variable is generated as an MA$(h-1)$ process:
\begin{equation}
  y_t = \mu + \sum_{k=0}^{h-1} \theta_k\, \varepsilon_{t-k}, \qquad \varepsilon_t \overset{\text{iid}}{\sim} N(0,1),
\label{eq:ucr}
\end{equation}
where $\theta_k = 0.5^k$ and $\mu$ is chosen so that the two models have equal expected squared-error loss. The resulting loss differential $d_{t+h} = e_{1,t+h}^2 - e_{2,t+h}^2$ is serially correlated up to displacement $R - 1$, which can far exceed $h - 1$. This DGP is challenging for tests that assume MA$(h-1)$ structure (DM-R, DM-M), because the loss differential's actual serial dependence extends well beyond the truncation lag.

\paragraph{Conditional-Rolling (CR) DGP.}
The target variable is generated as an ARMA process:
\begin{equation}
  y_t = \sum_{k=1}^{h-1} 0 \cdot y_{t-k} + \sum_{k=h}^{h+R-1} \frac{1}{2R}\, y_{t-k} + \varepsilon_t + \sum_{k=1}^{h-1} 0.5^k\, \varepsilon_{t-k},
\label{eq:cr}
\end{equation}
with 10{,}000 burn-in observations to reach stationarity. Under this DGP, the loss differential is serially correlated only up to displacement $h - 1$. The CR design is therefore favorable to the rectangular-kernel methods (DM-R, DM-M), whose truncation at $h - 1$ lags correctly captures the dependence structure.

\subsection{Simulation Design}

For each DGP, we vary four parameters:
\begin{itemize}[nosep]
  \item Forecast horizon: $h \in \{1, 3, 12\}$;
  \item Serial correlation range: $R \in \{25, 75, 125, 175\}$;
  \item Rolling window size: $\widetilde{R} \in \{25, 75, 125, 175\}$;
  \item Evaluation sample size: $P \in \{25, 75, 125, 175, 1000\}$.
\end{itemize}
This yields $3 \times 4 \times 4 \times 5 = 240$ parameter combinations per DGP. Each combination is replicated 5{,}000 times. At each replication, we apply nine tests at the 5\% nominal level: DM-R, DM-M, DM-NW, DM-NW-L, DM-FB, DM-EWC, and DM-IM with $q \in \{2, 5, 10\}$.
DM-NW uses the \citet{NeweyWest1994} plug-in bandwidth $M = \lceil 4(P/100)^{2/9}\rceil$ with normal critical values.
DM-NW-L uses the larger LLSW bandwidth $M = \lceil 1.3\sqrt{P}\rceil$ but retains normal critical values, isolating the effect of the bandwidth increase from the change in critical values.
For the fixed-smoothing procedures DM-FB and DM-EWC, we use the package default bandwidth choices: DM-FB sets $M = \lceil 1.3\sqrt{P}\rceil$ and DM-EWC sets $B = \lfloor 0.4 P^{2/3}\rfloor$, both following \citet{LLSW2018}.

The parameters $R$ and $\widetilde{R}$ play distinct roles. The serial correlation range $R$ is a feature of the DGP that determines the extent of dependence in the loss differential: under the UCR design, the loss differential is autocorrelated up to displacement $R - 1$, which can far exceed the forecast horizon $h$. The rolling window $\widetilde{R}$ is the in-sample estimation window used by the rolling-mean forecaster. When $R = \widetilde{R}$, the DGP is calibrated so that the two forecasters have equal expected loss---i.e., the null hypothesis holds exactly. When $R \neq \widetilde{R}$, the null is violated and one forecaster dominates, providing a setting for measuring power.

The ratio $P / R$ is also informative. When $P$ is small relative to $R$---as often arises in macroeconomic forecast evaluation, where evaluation samples are short---the serial correlation in the loss differential is large relative to the sample, and standard tests are most vulnerable to size distortion. Conversely, when $P$ is large (e.g., $P = 1{,}000$), all methods tend to be well-sized because the law of large numbers dominates.

\subsection{Size}

The \emph{empirical size} of a test is the fraction of the 5{,}000 replications in which the null hypothesis is rejected when $R = \widetilde{R}$ (i.e., under $H_0$). A well-calibrated test at the 5\% nominal level should yield empirical rejection frequencies close to 0.05.

Table~\ref{tab:size_summary} reports empirical size by forecast horizon $h$ for two evaluation sample sizes, $P \in \{75, 1{,}000\}$, in both DGPs. Each entry averages over the four on-diagonal values $R = \widetilde{R} \in \{25, 75, 125, 175\}$.

Under the UCR DGP (Panel~A), size distortion is concentrated in the smaller evaluation sample. The standard procedures DM-R, DM-M, and DM-NW over-reject, with the problem most pronounced at longer horizons. For example, at $P = 75$ and $h = 12$, DM-R rejects 0.160 of the time and DM-NW rejects 0.127, far above the nominal 0.05. By contrast, the fixed-smoothing methods DM-FB and DM-EWC remain close to nominal size across horizons and sample sizes. The IM test is well sized for $q = 2$ and becomes mildly oversized as the number of blocks increases.

The DM-NW-L row isolates the effect of simply increasing the bandwidth without adjusting the critical values. At $P = 75$, DM-NW-L over-rejects at rates of 0.099--0.118, \emph{worse} than DM-NW at short horizons ($h = 1, 3$) despite using the same LLSW bandwidth as DM-FB. The large bandwidth reduces bias in the long-run variance estimate but introduces substantial estimation noise; normal critical values, which do not account for this noise, lead to even larger size distortions. By contrast, DM-FB pairs the same bandwidth with fixed-$b$ critical values that widen the rejection region to account for estimation uncertainty, yielding size close to 0.05. At $P = 1{,}000$, DM-NW-L approaches the nominal level (0.050--0.052) as the estimation noise diminishes, whereas DM-FB is mildly conservative (0.038--0.040). This comparison demonstrates that the improved size control of the \citet{LLSW2018} recommendation comes from the combination of a large bandwidth \emph{and} fixed-$b$ critical values, not from the bandwidth increase alone.

Under the CR DGP (Panel~B), where the dependence structure is favorable to the rectangular-kernel truncation at $h-1$, size distortions are smaller but the same message holds in short samples. At $P = 75$ and $h = 12$, DM-R and DM-NW still reject at rates of 0.155 and 0.116, while fixed-smoothing procedures are much closer to 0.05. DM-NW-L again over-rejects (0.110--0.119), confirming that the large bandwidth alone does not deliver reliable size control. For DM-IM, larger block counts lead to over-rejection in the large sample, for example DM-IM($q{=}5$) rejects 0.093 at $P = 1{,}000$ and $h = 3$.

\begin{table}[t]
\centering
\caption{Summary of empirical size (nominal 5\%), by evaluation sample size $P$}
\label{tab:size_summary}
\small
\setlength{\tabcolsep}{4pt}
\begin{tabular}{@{}lccc|ccc@{}}
\toprule
& \multicolumn{3}{c|}{$P = 75$} & \multicolumn{3}{c}{$P = 1{,}000$} \\
\cmidrule(lr){2-4} \cmidrule(lr){5-7}
Method & $h=1$ & $h=3$ & $h=12$ & $h=1$ & $h=3$ & $h=12$ \\
\midrule
\multicolumn{7}{@{}l}{\textit{Panel A: Unconditional-Rolling (UCR) DGP}} \\[3pt]
DM-R & 0.047 & 0.061 & 0.160 & 0.025 & 0.025 & 0.040 \\
DM-M & 0.042 & 0.048 & 0.108 & 0.024 & 0.025 & 0.037 \\
DM-NW & 0.059 & 0.087 & 0.127 & 0.029 & 0.036 & 0.055 \\
DM-NW-L & 0.099 & 0.106 & 0.118 & 0.051 & 0.052 & 0.050 \\
DM-FB & 0.044 & 0.045 & 0.052 & 0.040 & 0.040 & 0.038 \\
DM-EWC & 0.052 & 0.041 & 0.045 & 0.038 & 0.037 & 0.036 \\
DM-IM ($q{=}2$) & 0.051 & 0.047 & 0.046 & 0.052 & 0.052 & 0.048 \\
DM-IM ($q{=}5$) & 0.055 & 0.047 & 0.045 & 0.067 & 0.066 & 0.057 \\
DM-IM ($q{=}10$) & 0.052 & 0.052 & 0.062 & 0.061 & 0.062 & 0.053 \\
\midrule
\multicolumn{7}{@{}l}{\textit{Panel B: Conditional-Rolling (CR) DGP}} \\[3pt]
DM-R & 0.055 & 0.068 & 0.155 & 0.049 & 0.052 & 0.055 \\
DM-M & 0.049 & 0.054 & 0.103 & 0.048 & 0.052 & 0.053 \\
DM-NW & 0.070 & 0.096 & 0.116 & 0.055 & 0.068 & 0.074 \\
DM-NW-L & 0.113 & 0.119 & 0.110 & 0.082 & 0.082 & 0.068 \\
DM-FB & 0.053 & 0.051 & 0.048 & 0.068 & 0.067 & 0.053 \\
DM-EWC & 0.060 & 0.046 & 0.043 & 0.064 & 0.063 & 0.049 \\
DM-IM ($q{=}2$) & 0.052 & 0.049 & 0.044 & 0.058 & 0.060 & 0.055 \\
DM-IM ($q{=}5$) & 0.062 & 0.052 & 0.041 & 0.096 & 0.093 & 0.072 \\
DM-IM ($q{=}10$) & 0.061 & 0.058 & 0.058 & 0.092 & 0.087 & 0.070 \\
\bottomrule
\end{tabular}
\begin{flushleft}
\footnotesize
\textit{Notes:} Each entry is the mean rejection frequency across on-diagonal designs $R = \widetilde{R} \in \{25, 75, 125, 175\}$ for the given $(P, h)$ pair. Reported panels are restricted to $P \in \{75, 1{,}000\}$. Complete cell-by-cell results are in Appendix~\ref{sec:appendix_tables}.
\end{flushleft}
\end{table}

\subsection{Size-Corrected Power}

Raw rejection frequencies under $H_1$ (i.e., when $\widetilde{R} \neq R$) confound the test's genuine ability to detect deviations from the null with its tendency to over-reject under $H_0$. To isolate the former, we compute \emph{size-corrected power}. For each on-diagonal configuration $(R = \widetilde{R}, P, h)$, we extract the 95th percentile of $|\text{DM}|$ across the 5{,}000 replications as the size-corrected critical value:
\begin{equation}
  c^*_{0.95}(R, P, h) = Q_{0.95}\!\left(\left\{|\text{DM}_i|\right\}_{i=1}^{5000}\right).
\end{equation}
We then recompute the rejection frequency for all off-diagonal configurations ($\widetilde{R} \neq R$) using this critical value instead of the asymptotic one:
\begin{equation}
  \text{Size-corrected power} = \frac{1}{5000}\sum_{i=1}^{5000} \boldsymbol{1}\!\left\{|\text{DM}_i| > c^*_{0.95}(R, P, h)\right\}.
\end{equation}
Because the critical value $c^*_{0.95}$ is calibrated to deliver exactly 5\% rejection under $H_0$, the size-corrected power strips away size distortion and measures only the test's ability to discriminate between the null and alternative.

Table~\ref{tab:power_summary} reports the size-corrected power under the UCR DGP for two representative scenarios that hold $P$ and $\widetilde{R}$ fixed. In each panel, the rolling window is fixed at $\widetilde{R} = 25$ and the power is averaged over the three off-diagonal values $R \in \{75, 125, 175\}$. Panel~A uses $P = 75$ (a small evaluation sample typical of macroeconomic forecast comparisons), while Panel~B uses $P = 1{,}000$ (a large sample in which all tests have substantial power). We report DM-IM with $q \in \{2, 5, 10\}$ to examine how the number of blocks affects the size--power trade-off.

\begin{table}[t]
\centering
\caption{Size-corrected power, UCR DGP ($\widetilde{R} = 25$, averaged over $R \in \{75, 125, 175\}$)}
\label{tab:power_summary}
\small
\begin{tabular}{@{}lccc|ccc@{}}
\toprule
& \multicolumn{3}{c|}{Panel A: $P = 75$} & \multicolumn{3}{c}{Panel B: $P = 1{,}000$} \\
\cmidrule(lr){2-4} \cmidrule(lr){5-7}
Method & $h=1$ & $h=3$ & $h=12$ & $h=1$ & $h=3$ & $h=12$ \\
\midrule
DM-R & 0.086 & 0.078 & 0.066 & 0.751 & 0.710 & 0.606 \\
DM-M & 0.086 & 0.078 & 0.066 & 0.751 & 0.710 & 0.606 \\
DM-NW & 0.104 & 0.079 & 0.076 & 0.766 & 0.715 & 0.628 \\
DM-NW-L & 0.127 & 0.097 & 0.068 & 0.806 & 0.754 & 0.627 \\
DM-FB & 0.127 & 0.097 & 0.068 & 0.806 & 0.754 & 0.627 \\
DM-EWC & 0.121 & 0.092 & 0.062 & 0.812 & 0.756 & 0.630 \\
DM-IM ($q{=}2$) & 0.085 & 0.077 & 0.054 & 0.194 & 0.180 & 0.133 \\
DM-IM ($q{=}5$) & 0.124 & 0.106 & 0.066 & 0.568 & 0.500 & 0.339 \\
DM-IM ($q{=}10$) & 0.119 & 0.087 & 0.070 & 0.709 & 0.648 & 0.493 \\
\bottomrule
\end{tabular}
\begin{flushleft}
\footnotesize
\textit{Notes:} Each entry is the mean size-corrected rejection frequency across $R \in \{75, 125, 175\}$ for the given $(P, h)$ pair and $\widetilde{R} = 25$. Complete cell-by-cell results are in Appendix~\ref{sec:appendix_tables}.
\end{flushleft}
\end{table}

Table~\ref{tab:power_summary} shows that, after correcting for size, fixed-smoothing methods do not sacrifice power relative to traditional normal-approximation procedures. In the small-sample case (Panel~A, $P = 75$), the fixed-$b$ Bartlett test (DM-FB) and the EWC-based test (DM-EWC) deliver size-corrected power that is comparable to, and often higher than, DM-R and DM-NW (for example, at $h = 1$, 0.127 and 0.121 versus 0.086 and 0.104). In the large-sample case (Panel~B, $P = 1{,}000$), all kernel-based methods have high power, with DM-FB and DM-EWC matching or slightly exceeding DM-R. Notably, DM-NW-L and DM-FB have identical size-corrected power in every cell, which is expected: both use the same bandwidth and hence the same test statistic, so size correction equalizes them. The difference between the two lies entirely in the choice of critical values, which affects only uncorrected size. Thus, in our designs, the improved size control of fixed-smoothing inference does not come at a meaningful power cost. As expected, the IM test gains power as the number of blocks increases, and power declines at longer horizons.

\section{Conclusion}
\label{sec:conclusion}

The \pkg{ForeComp} package provides an \R{} toolkit for comparing predictive accuracy using Diebold--Mariano type tests under a common interface. It includes both standard and fixed-smoothing procedures and the \fn{Plot\_Tradeoff} diagnostic for bandwidth sensitivity. Across three Survey of Professional Forecasters applications and Monte Carlo experiments, we find that standard normal-approximation tests can substantially over-reject in short samples, while fixed-smoothing methods offer improved size control with competitive size-corrected power. We recommend reporting fixed-smoothing results and using \fn{Plot\_Tradeoff} to assess bandwidth robustness. 


\bibliographystyle{apalike}

\clearpage
\appendix

\section{Additional Replication Tables for Application~2}
\label{sec:appendix_ci}

This appendix reports the remaining tables from the replication of \citet{CoroneoIacone2020}. Table~\ref{tab:ci_table1} (real output growth) appears in the main text. Tables~\ref{tab:ci_table2}--\ref{tab:ci_table5} below complete the replication of all five tables from the original paper.

\begin{table}[ht]
\centering
\caption{GNP/GDP inflation: SPF versus random walk (replication of CI Table~2)}
\label{tab:ci_table2}
\small
\begin{tabular}{@{}lccccc@{}}
\toprule
& \multicolumn{5}{c}{Forecast horizon} \\
\cmidrule(l){2-6}
& $h=0$ & $h=1$ & $h=2$ & $h=3$ & $h=4$ \\
\midrule
\multicolumn{6}{@{}l}{\textit{Evaluation period: 1987:Q1--2016:Q4, $T = 120$}} \\[3pt]
WCE-DM  & 3.99 & 3.87 & 2.52 & 1.18 & 2.46 \\
WCE-B   & 2.81** & 3.68** & 2.72** & 1.02 & 2.54** \\
WPE-D   & 2.48** & 3.53** & 2.38** & 0.86 & 2.25* \\[6pt]
\multicolumn{6}{@{}l}{\textit{Evaluation period: 1987:Q1--1996:Q4, $T = 40$}} \\[3pt]
WCE-DM  & 1.31 & 2.42 & 1.22 & 0.04 & 1.31 \\
WCE-B   & 1.36 & 2.37* & 1.39 & 0.03 & 1.38 \\
WPE-D   & 1.40 & 2.09* & 1.55 & 0.03 & 1.17 \\[6pt]
\multicolumn{6}{@{}l}{\textit{Evaluation period: 1997:Q1--2006:Q4, $T = 40$}} \\[3pt]
WCE-DM  & 2.28 & 1.67 & 1.06 & 0.98 & 0.67 \\
WCE-B   & 2.23* & 1.66 & 0.99 & 1.01 & 0.78 \\
WPE-D   & 2.25* & 1.87 & 0.97 & 0.98 & 0.72 \\[6pt]
\multicolumn{6}{@{}l}{\textit{Evaluation period: 2007:Q1--2016:Q4, $T = 40$}} \\[3pt]
WCE-DM  & 3.66 & 3.17 & 2.10 & 0.97 & 3.13 \\
WCE-B   & 2.75** & 2.98** & 2.29* & 0.93 & 2.31* \\
WPE-D   & 2.47** & 3.10** & 1.91 & 0.86 & 2.25* \\
\bottomrule
\end{tabular}
\begin{flushleft}
\footnotesize
\textit{Notes:} WCE-DM uses a normal approximation, with two-sided 5\% and 10\% critical values 1.96 and 1.645. WCE-B (fixed-$b$) and WPE-D (fixed-$m$) use fixed-smoothing asymptotics and associated reference distributions rather than the standard normal approximation; see Section~\ref{sec:fixedsmoothing} for details. Asterisks are reported only for WCE-B and WPE-D, where ** and * denote two-sided significance at the 5\% and 10\% levels.
\end{flushleft}
\end{table}

\begin{table}[ht]
\centering
\caption{Unemployment rate: SPF versus random walk (replication of CI Table~3)}
\label{tab:ci_table3}
\small
\begin{tabular}{@{}lccccc@{}}
\toprule
& \multicolumn{5}{c}{Forecast horizon} \\
\cmidrule(l){2-6}
& $h=0$ & $h=1$ & $h=2$ & $h=3$ & $h=4$ \\
\midrule
\multicolumn{6}{@{}l}{\textit{Evaluation period: 1987:Q1--2016:Q4, $T = 120$}} \\[3pt]
WCE-DM  & 3.84 & 2.15 & 1.98 & 2.06 & 2.27 \\
WCE-B   & 2.37** & 1.99* & 2.05* & 2.20* & 2.46** \\
WPE-D   & 2.14* & 1.86* & 1.88* & 2.00* & 2.20* \\[6pt]
\multicolumn{6}{@{}l}{\textit{Evaluation period: 1987:Q1--1996:Q4, $T = 40$}} \\[3pt]
WCE-DM  & 3.47 & 1.72 & 1.66 & 2.07 & 2.52 \\
WCE-B   & 3.13** & 1.89 & 1.93 & 2.37* & 2.57** \\
WPE-D   & 2.53** & 1.61 & 1.58 & 1.89 & 2.01* \\[6pt]
\multicolumn{6}{@{}l}{\textit{Evaluation period: 1997:Q1--2006:Q4, $T = 40$}} \\[3pt]
WCE-DM  & 2.17 & 1.75 & 1.47 & 1.23 & 1.15 \\
WCE-B   & 2.23* & 1.75 & 1.60 & 1.40 & 1.31 \\
WPE-D   & 2.02* & 1.51 & 1.36 & 1.17 & 1.09 \\[6pt]
\multicolumn{6}{@{}l}{\textit{Evaluation period: 2007:Q1--2016:Q4, $T = 40$}} \\[3pt]
WCE-DM  & 2.83 & 1.72 & 1.64 & 1.82 & 2.19 \\
WCE-B   & 1.81 & 1.68 & 1.80 & 2.04* & 2.42** \\
WPE-D   & 1.57 & 1.43 & 1.52 & 1.73 & 2.04* \\
\bottomrule
\end{tabular}
\begin{flushleft}
\footnotesize
\textit{Notes:} WCE-DM uses a normal approximation, with two-sided 5\% and 10\% critical values 1.96 and 1.645. WCE-B (fixed-$b$) and WPE-D (fixed-$m$) use fixed-smoothing asymptotics and associated reference distributions rather than the standard normal approximation; see Section~\ref{sec:fixedsmoothing} for details. Asterisks are reported only for WCE-B and WPE-D, where ** and * denote two-sided significance at the 5\% and 10\% levels.
\end{flushleft}
\end{table}

\begin{table}[ht]
\centering
\caption{Three-month Treasury bill: SPF versus random walk (replication of CI Table~4)}
\label{tab:ci_table4}
\small
\begin{tabular}{@{}lccccc@{}}
\toprule
& \multicolumn{5}{c}{Forecast horizon} \\
\cmidrule(l){2-6}
& $h=0$ & $h=1$ & $h=2$ & $h=3$ & $h=4$ \\
\midrule
\multicolumn{6}{@{}l}{\textit{Evaluation period: 1987:Q1--2016:Q4, $T = 120$}} \\[3pt]
WCE-DM  & 5.05 & 3.79 & 3.16 & 2.72 & 1.76 \\
WCE-B   & 3.84** & 3.78** & 3.54** & 3.11** & 1.96* \\
WPE-D   & 3.49** & 3.58** & 3.52** & 3.27** & 2.00* \\[6pt]
\multicolumn{6}{@{}l}{\textit{Evaluation period: 1987:Q1--1996:Q4, $T = 40$}} \\[3pt]
WCE-DM  & 4.73 & 3.01 & 2.58 & 2.77 & 2.82 \\
WCE-B   & 4.10** & 3.50** & 3.10** & 2.94** & 2.63** \\
WPE-D   & 3.17** & 2.68** & 2.56** & 2.53** & 2.10* \\[6pt]
\multicolumn{6}{@{}l}{\textit{Evaluation period: 1997:Q1--2006:Q4, $T = 40$}} \\[3pt]
WCE-DM  & 3.02 & 2.46 & 2.10 & 1.69 & 1.07 \\
WCE-B   & 2.18* & 2.34* & 2.28* & 1.85 & 1.16 \\
WPE-D   & 1.82 & 1.93 & 1.87 & 1.52 & 0.95 \\[6pt]
\multicolumn{6}{@{}l}{\textit{Evaluation period: 2007:Q1--2016:Q4, $T = 40$}} \\[3pt]
WCE-DM  & 1.68 & 1.19 & 1.14 & 0.77 & $-$0.47 \\
WCE-B   & 1.33 & 1.23 & 1.03 & 0.71 & $-$0.46 \\
WPE-D   & 1.17 & 1.09 & 0.92 & 0.65 & $-$0.45 \\
\bottomrule
\end{tabular}
\begin{flushleft}
\footnotesize
\textit{Notes:} WCE-DM uses a normal approximation, with two-sided 5\% and 10\% critical values 1.96 and 1.645. WCE-B (fixed-$b$) and WPE-D (fixed-$m$) use fixed-smoothing asymptotics and associated reference distributions rather than the standard normal approximation; see Section~\ref{sec:fixedsmoothing} for details. Asterisks are reported only for WCE-B and WPE-D, where ** and * denote two-sided significance at the 5\% and 10\% levels.
\end{flushleft}
\end{table}

\begin{table}[ht]
\centering
\caption{ECB Survey of Professional Forecasters (replication of CI Table~5)}
\label{tab:ci_table5}
\small
\begin{tabular}{@{}lccc|ccc@{}}
\toprule
& \multicolumn{3}{c|}{GDP growth} & \multicolumn{3}{c}{HICP inflation} \\
\cmidrule(lr){2-4} \cmidrule(l){5-7}
Forecast horizon (years) & 1 & 2 & 5 & 1 & 2 & 5 \\
\midrule
WCE-DM  & 2.09 & 1.80 & 1.04 & 1.28 & 1.75 & 1.04 \\
WCE-B   & 1.84 & 2.04* & 1.08 & 1.18 & 1.79 & 0.89 \\
WPE-D   & 1.54 & 1.81 & 0.89 & 0.97 & 1.48 & 0.71 \\
\bottomrule
\end{tabular}
\begin{flushleft}
\footnotesize
\textit{Notes:} Evaluation period: 2006:Q1--2016:Q4 ($T = 44$). WCE-DM uses a normal approximation, with two-sided 5\% and 10\% critical values 1.96 and 1.645. WCE-B (fixed-$b$) and WPE-D (fixed-$m$) use fixed-smoothing asymptotics and associated reference distributions rather than the standard normal approximation; see Section~\ref{sec:fixedsmoothing} for details. Asterisks are reported only for WCE-B and WPE-D, where ** and * denote two-sided significance at the 5\% and 10\% levels.
\end{flushleft}
\end{table}

\clearpage
\section{Monte Carlo Simulation Tables}
\label{sec:appendix_tables}

This appendix reports the complete Monte Carlo results for all nine tests considered in the simulation design under both DGPs. Each table is a $16 \times 15$ matrix with rows indexed by $(R, \widetilde{R})$ and columns indexed by $(h, P)$. Diagonal entries ($R = \widetilde{R}$, shown in bold) give the empirical size; off-diagonal entries give (raw or size-corrected) rejection frequencies.

For DM-IM, if $n=P$ is not divisible by the chosen number of blocks $q$, we use a balanced nonoverlapping partition: letting $b_0=\lfloor n/q \rfloor$ and $r=n \bmod q$, the first $r$ blocks have size $b_0+1$ and the remaining $q-r$ blocks have size $b_0$. Hence block sizes differ by at most one.

For ease of reference, the tables are grouped into four blocks:
\begin{itemize}[nosep]
  \item \textbf{UCR Size Tables}: empirical size under the Unconditional-Rolling (UCR) DGP.
  \item \textbf{UCR Size-Corrected Power Tables}: size-corrected power under the UCR DGP.
  \item \textbf{CR Size Tables}: empirical size under the Conditional-Rolling (CR) DGP.
  \item \textbf{CR Size-Corrected Power Tables}: size-corrected power under the CR DGP.
\end{itemize}

\subsection*{UCR Size Tables}
\begin{table}[H]
\centering
\caption{Size, Unconditional-Rolling DGP, DM-R}
\scalebox{0.7}{
\begin{tabular}{cc|ccccc|ccccc|ccccc}
\toprule
\toprule
&       & \multicolumn{5}{c|}{$h=1$}            & \multicolumn{5}{c|}{$h=3$}            & \multicolumn{5}{c}{$h=12$} \\
\cmidrule{3-17}          &       & \multicolumn{5}{c|}{$P$}              & \multicolumn{5}{c|}{$P$}              & \multicolumn{5}{c}{$P$} \\
$R$   & $\widetilde{R}$ & 25    & 75    & 125   & 175   & 1000  & 25    & 75    & 125   & 175   & 1000  & 25    & 75    & 125   & 175   & 1000 \\
\midrule
25 & 25& \textbf{0.058}& \textbf{0.034}& \textbf{0.030}& \textbf{0.026}& \textbf{0.019}& \textbf{0.109}& \textbf{0.051}& \textbf{0.044}& \textbf{0.040}& \textbf{0.026}& \textbf{0.349}& \textbf{0.166}& \textbf{0.127}& \textbf{0.106}& \textbf{0.049}\\
25 & 75& 0.070& 0.061& 0.045& 0.048& 0.406& 0.142& 0.072& 0.065& 0.071& 0.395& 0.374& 0.212& 0.173& 0.160& 0.410\\
25 & 125& 0.075& 0.081& 0.092& 0.096& 0.670& 0.160& 0.109& 0.111& 0.119& 0.657& 0.386& 0.243& 0.214& 0.207& 0.623\\
25 & 175& 0.083& 0.101& 0.124& 0.142& 0.783& 0.154& 0.127& 0.149& 0.164& 0.763& 0.369& 0.262& 0.250& 0.242& 0.746\\
\midrule
75 & 25& 0.061& 0.078& 0.099& 0.115& 0.466& 0.103& 0.084& 0.097& 0.117& 0.426& 0.354& 0.187& 0.145& 0.158& 0.386\\
\bottomrule
\bottomrule
\end{tabular}
}
\label{tab:UCR_size_DM-R}
\end{table}

\begin{table}[H]
\centering
\caption{Size, Unconditional-Rolling DGP, DM-M}
\scalebox{0.7}{
\begin{tabular}{cc|ccccc|ccccc|ccccc}
\toprule
\toprule
&       & \multicolumn{5}{c|}{$h=1$}            & \multicolumn{5}{c|}{$h=3$}            & \multicolumn{5}{c}{$h=12$} \\
\cmidrule{3-17}          &       & \multicolumn{5}{c|}{$P$}              & \multicolumn{5}{c|}{$P$}              & \multicolumn{5}{c}{$P$} \\
$R$   & $\widetilde{R}$ & 25    & 75    & 125   & 175   & 1000  & 25    & 75    & 125   & 175   & 1000  & 25    & 75    & 125   & 175   & 1000 \\
\midrule
25 & 25& \textbf{0.040}& \textbf{0.031}& \textbf{0.028}& \textbf{0.025}& \textbf{0.019}& \textbf{0.063}& \textbf{0.041}& \textbf{0.038}& \textbf{0.035}& \textbf{0.026}& \textbf{0.129}& \textbf{0.115}& \textbf{0.094}& \textbf{0.083}& \textbf{0.046}\\
25 & 75& 0.053& 0.054& 0.042& 0.046& 0.404& 0.093& 0.058& 0.056& 0.062& 0.393& 0.157& 0.148& 0.136& 0.133& 0.397\\
25 & 125& 0.060& 0.074& 0.086& 0.092& 0.669& 0.110& 0.094& 0.102& 0.111& 0.653& 0.167& 0.176& 0.170& 0.175& 0.609\\
25 & 175& 0.063& 0.093& 0.118& 0.136& 0.782& 0.105& 0.110& 0.136& 0.151& 0.759& 0.160& 0.197& 0.211& 0.207& 0.737\\
\midrule
75 & 25& 0.045& 0.070& 0.095& 0.112& 0.464& 0.056& 0.068& 0.085& 0.109& 0.422& 0.122& 0.124& 0.106& 0.124& 0.375\\
\bottomrule
\bottomrule
\end{tabular}
}
\label{tab:UCR_size_DM-M}
\end{table}

\begin{table}[H]
\centering
\caption{Size, Unconditional-Rolling DGP, DM-NW}
\scalebox{0.7}{
\begin{tabular}{cc|ccccc|ccccc|ccccc}
\toprule
\toprule
&       & \multicolumn{5}{c|}{$h=1$}            & \multicolumn{5}{c|}{$h=3$}            & \multicolumn{5}{c}{$h=12$} \\
\cmidrule{3-17}          &       & \multicolumn{5}{c|}{$P$}              & \multicolumn{5}{c|}{$P$}              & \multicolumn{5}{c}{$P$} \\
$R$   & $\widetilde{R}$ & 25    & 75    & 125   & 175   & 1000  & 25    & 75    & 125   & 175   & 1000  & 25    & 75    & 125   & 175   & 1000 \\
\midrule
25 & 25& \textbf{0.084}& \textbf{0.046}& \textbf{0.042}& \textbf{0.040}& \textbf{0.026}& \textbf{0.126}& \textbf{0.072}& \textbf{0.062}& \textbf{0.058}& \textbf{0.037}& \textbf{0.203}& \textbf{0.128}& \textbf{0.095}& \textbf{0.100}& \textbf{0.071}\\
25 & 75& 0.090& 0.072& 0.058& 0.060& 0.418& 0.159& 0.102& 0.093& 0.097& 0.457& 0.242& 0.168& 0.140& 0.141& 0.480\\
25 & 125& 0.099& 0.097& 0.106& 0.110& 0.678& 0.184& 0.144& 0.146& 0.161& 0.707& 0.246& 0.198& 0.193& 0.203& 0.720\\
25 & 175& 0.103& 0.119& 0.140& 0.154& 0.782& 0.177& 0.169& 0.183& 0.211& 0.807& 0.239& 0.222& 0.233& 0.250& 0.821\\
\midrule
75 & 25& 0.099& 0.114& 0.138& 0.157& 0.518& 0.129& 0.121& 0.139& 0.162& 0.491& 0.195& 0.166& 0.156& 0.188& 0.477\\
\bottomrule
\bottomrule
\end{tabular}
}
\label{tab:UCR_size_DM-NW}
\end{table}

\begin{table}[H]
\centering
\caption{Size, Unconditional-Rolling DGP, DM-NW-L}
\scalebox{0.7}{
\begin{tabular}{cc|ccccc|ccccc|ccccc}
\toprule
\toprule
&       & \multicolumn{5}{c|}{$h=1$}            & \multicolumn{5}{c|}{$h=3$}            & \multicolumn{5}{c}{$h=12$} \\
\cmidrule{3-17}          &       & \multicolumn{5}{c|}{$P$}              & \multicolumn{5}{c|}{$P$}              & \multicolumn{5}{c}{$P$} \\
$R$   & $\widetilde{R}$ & 25    & 75    & 125   & 175   & 1000  & 25    & 75    & 125   & 175   & 1000  & 25    & 75    & 125   & 175   & 1000 \\
\midrule
25 & 25& \textbf{0.133}& \textbf{0.094}& \textbf{0.079}& \textbf{0.079}& \textbf{0.056}& \textbf{0.155}& \textbf{0.096}& \textbf{0.087}& \textbf{0.082}& \textbf{0.061}& \textbf{0.211}& \textbf{0.124}& \textbf{0.094}& \textbf{0.094}& \textbf{0.062}\\
25 & 75& 0.136& 0.106& 0.092& 0.089& 0.467& 0.188& 0.117& 0.110& 0.118& 0.471& 0.249& 0.164& 0.141& 0.145& 0.448\\
25 & 125& 0.164& 0.141& 0.142& 0.150& 0.702& 0.211& 0.157& 0.167& 0.174& 0.697& 0.253& 0.193& 0.188& 0.196& 0.659\\
25 & 175& 0.163& 0.156& 0.179& 0.194& 0.794& 0.219& 0.185& 0.197& 0.221& 0.787& 0.250& 0.219& 0.231& 0.235& 0.772\\
\midrule
75 & 25& 0.158& 0.190& 0.222& 0.260& 0.666& 0.151& 0.169& 0.188& 0.222& 0.598& 0.204& 0.142& 0.128& 0.157& 0.454\\
\bottomrule
\bottomrule
\end{tabular}
}
\label{tab:UCR_size_DM-NW-L}
\end{table}

\begin{table}[H]
\centering
\caption{Size, Unconditional-Rolling DGP, DM-FB}
\scalebox{0.7}{
\begin{tabular}{cc|ccccc|ccccc|ccccc}
\toprule
\toprule
&       & \multicolumn{5}{c|}{$h=1$}            & \multicolumn{5}{c|}{$h=3$}            & \multicolumn{5}{c}{$h=12$} \\
\cmidrule{3-17}          &       & \multicolumn{5}{c|}{$P$}              & \multicolumn{5}{c|}{$P$}              & \multicolumn{5}{c}{$P$} \\
$R$   & $\widetilde{R}$ & 25    & 75    & 125   & 175   & 1000  & 25    & 75    & 125   & 175   & 1000  & 25    & 75    & 125   & 175   & 1000 \\
\midrule
25 & 25& \textbf{0.038}& \textbf{0.043}& \textbf{0.044}& \textbf{0.048}& \textbf{0.043}& \textbf{0.048}& \textbf{0.042}& \textbf{0.049}& \textbf{0.046}& \textbf{0.046}& \textbf{0.077}& \textbf{0.059}& \textbf{0.049}& \textbf{0.055}& \textbf{0.046}\\
25 & 75& 0.042& 0.045& 0.044& 0.046& 0.413& 0.074& 0.053& 0.057& 0.066& 0.410& 0.117& 0.084& 0.078& 0.088& 0.394\\
25 & 125& 0.056& 0.068& 0.076& 0.091& 0.648& 0.093& 0.085& 0.092& 0.106& 0.638& 0.122& 0.118& 0.119& 0.124& 0.600\\
25 & 175& 0.056& 0.080& 0.105& 0.124& 0.742& 0.094& 0.102& 0.126& 0.140& 0.735& 0.121& 0.135& 0.151& 0.160& 0.723\\
\midrule
75 & 25& 0.048& 0.105& 0.147& 0.188& 0.625& 0.045& 0.080& 0.111& 0.151& 0.557& 0.068& 0.064& 0.068& 0.096& 0.404\\
\bottomrule
\bottomrule
\end{tabular}
}
\label{tab:UCR_size_DM-FB}
\end{table}

\begin{table}[H]
\centering
\caption{Size, Unconditional-Rolling DGP, DM-EWC}
\scalebox{0.7}{
\begin{tabular}{cc|ccccc|ccccc|ccccc}
\toprule
\toprule
&       & \multicolumn{5}{c|}{$h=1$}            & \multicolumn{5}{c|}{$h=3$}            & \multicolumn{5}{c}{$h=12$} \\
\cmidrule{3-17}          &       & \multicolumn{5}{c|}{$P$}              & \multicolumn{5}{c|}{$P$}              & \multicolumn{5}{c}{$P$} \\
$R$   & $\widetilde{R}$ & 25    & 75    & 125   & 175   & 1000  & 25    & 75    & 125   & 175   & 1000  & 25    & 75    & 125   & 175   & 1000 \\
\midrule
25 & 25& \textbf{0.046}& \textbf{0.052}& \textbf{0.050}& \textbf{0.054}& \textbf{0.043}& \textbf{0.036}& \textbf{0.039}& \textbf{0.044}& \textbf{0.044}& \textbf{0.045}& \textbf{0.042}& \textbf{0.053}& \textbf{0.045}& \textbf{0.052}& \textbf{0.048}\\
25 & 75& 0.048& 0.052& 0.049& 0.051& 0.410& 0.050& 0.048& 0.053& 0.061& 0.400& 0.068& 0.069& 0.071& 0.080& 0.379\\
25 & 125& 0.056& 0.072& 0.081& 0.094& 0.651& 0.063& 0.076& 0.088& 0.099& 0.634& 0.080& 0.097& 0.106& 0.116& 0.591\\
25 & 175& 0.055& 0.083& 0.110& 0.129& 0.751& 0.068& 0.090& 0.118& 0.133& 0.729& 0.074& 0.114& 0.139& 0.152& 0.719\\
\midrule
75 & 25& 0.054& 0.111& 0.147& 0.194& 0.625& 0.032& 0.070& 0.100& 0.136& 0.540& 0.036& 0.052& 0.055& 0.082& 0.386\\
\bottomrule
\bottomrule
\end{tabular}
}
\label{tab:UCR_size_DM-EWC}
\end{table}

\begin{table}[H]
\centering
\caption{Size, Unconditional-Rolling DGP, DM-IM2}
\scalebox{0.7}{
\begin{tabular}{cc|ccccc|ccccc|ccccc}
\toprule
\toprule
&       & \multicolumn{5}{c|}{$h=1$}            & \multicolumn{5}{c|}{$h=3$}            & \multicolumn{5}{c}{$h=12$} \\
\cmidrule{3-17}          &       & \multicolumn{5}{c|}{$P$}              & \multicolumn{5}{c|}{$P$}              & \multicolumn{5}{c}{$P$} \\
$R$   & $\widetilde{R}$ & 25    & 75    & 125   & 175   & 1000  & 25    & 75    & 125   & 175   & 1000  & 25    & 75    & 125   & 175   & 1000 \\
\midrule
25 & 25& \textbf{0.043}& \textbf{0.058}& \textbf{0.055}& \textbf{0.048}& \textbf{0.049}& \textbf{0.050}& \textbf{0.051}& \textbf{0.051}& \textbf{0.050}& \textbf{0.053}& \textbf{0.046}& \textbf{0.045}& \textbf{0.043}& \textbf{0.056}& \textbf{0.045}\\
25 & 75& 0.048& 0.049& 0.050& 0.053& 0.121& 0.049& 0.051& 0.047& 0.059& 0.114& 0.058& 0.050& 0.057& 0.058& 0.110\\
25 & 125& 0.054& 0.049& 0.054& 0.067& 0.140& 0.053& 0.049& 0.055& 0.063& 0.150& 0.058& 0.057& 0.055& 0.063& 0.137\\
25 & 175& 0.055& 0.059& 0.067& 0.067& 0.153& 0.057& 0.059& 0.061& 0.066& 0.154& 0.059& 0.067& 0.063& 0.070& 0.146\\
\midrule
75 & 25& 0.044& 0.079& 0.087& 0.095& 0.162& 0.047& 0.063& 0.077& 0.080& 0.149& 0.049& 0.049& 0.060& 0.053& 0.105\\
\bottomrule
\bottomrule
\end{tabular}
}
\label{tab:UCR_size_DM-IM2}
\end{table}

\begin{table}[H]
\centering
\caption{Size, Unconditional-Rolling DGP, DM-IM5}
\scalebox{0.7}{
\begin{tabular}{cc|ccccc|ccccc|ccccc}
\toprule
\toprule
&       & \multicolumn{5}{c|}{$h=1$}            & \multicolumn{5}{c|}{$h=3$}            & \multicolumn{5}{c}{$h=12$} \\
\cmidrule{3-17}          &       & \multicolumn{5}{c|}{$P$}              & \multicolumn{5}{c|}{$P$}              & \multicolumn{5}{c}{$P$} \\
$R$   & $\widetilde{R}$ & 25    & 75    & 125   & 175   & 1000  & 25    & 75    & 125   & 175   & 1000  & 25    & 75    & 125   & 175   & 1000 \\
\midrule
25 & 25& \textbf{0.048}& \textbf{0.059}& \textbf{0.069}& \textbf{0.068}& \textbf{0.054}& \textbf{0.047}& \textbf{0.051}& \textbf{0.057}& \textbf{0.062}& \textbf{0.052}& \textbf{0.061}& \textbf{0.053}& \textbf{0.041}& \textbf{0.048}& \textbf{0.047}\\
25 & 75& 0.046& 0.056& 0.050& 0.061& 0.272& 0.064& 0.050& 0.058& 0.064& 0.269& 0.097& 0.072& 0.061& 0.074& 0.261\\
25 & 125& 0.053& 0.063& 0.076& 0.081& 0.427& 0.073& 0.070& 0.073& 0.090& 0.429& 0.105& 0.099& 0.094& 0.101& 0.405\\
25 & 175& 0.060& 0.071& 0.096& 0.101& 0.515& 0.084& 0.083& 0.099& 0.110& 0.491& 0.096& 0.105& 0.115& 0.121& 0.465\\
\midrule
75 & 25& 0.058& 0.116& 0.171& 0.226& 0.510& 0.042& 0.089& 0.125& 0.174& 0.440& 0.054& 0.055& 0.061& 0.076& 0.278\\
\bottomrule
\bottomrule
\end{tabular}
}
\label{tab:UCR_size_DM-IM5}
\end{table}

\begin{table}[H]
\centering
\caption{Size, Unconditional-Rolling DGP, DM-IM10}
\scalebox{0.7}{
\begin{tabular}{cc|ccccc|ccccc|ccccc}
\toprule
\toprule
&       & \multicolumn{5}{c|}{$h=1$}            & \multicolumn{5}{c|}{$h=3$}            & \multicolumn{5}{c}{$h=12$} \\
\cmidrule{3-17}          &       & \multicolumn{5}{c|}{$P$}              & \multicolumn{5}{c|}{$P$}              & \multicolumn{5}{c}{$P$} \\
$R$   & $\widetilde{R}$ & 25    & 75    & 125   & 175   & 1000  & 25    & 75    & 125   & 175   & 1000  & 25    & 75    & 125   & 175   & 1000 \\
\midrule
25 & 25& \textbf{0.048}& \textbf{0.048}& \textbf{0.056}& \textbf{0.058}& \textbf{0.052}& \textbf{0.073}& \textbf{0.047}& \textbf{0.054}& \textbf{0.055}& \textbf{0.056}& \textbf{0.135}& \textbf{0.066}& \textbf{0.051}& \textbf{0.056}& \textbf{0.050}\\
25 & 75& 0.052& 0.058& 0.052& 0.057& 0.379& 0.097& 0.058& 0.063& 0.070& 0.365& 0.173& 0.087& 0.079& 0.094& 0.348\\
25 & 125& 0.055& 0.069& 0.082& 0.092& 0.585& 0.124& 0.089& 0.095& 0.107& 0.562& 0.176& 0.118& 0.116& 0.124& 0.531\\
25 & 175& 0.057& 0.083& 0.109& 0.122& 0.670& 0.123& 0.103& 0.121& 0.136& 0.647& 0.166& 0.135& 0.152& 0.153& 0.631\\
\midrule
75 & 25& 0.054& 0.110& 0.161& 0.204& 0.607& 0.070& 0.087& 0.118& 0.161& 0.540& 0.127& 0.080& 0.074& 0.091& 0.365\\
\bottomrule
\bottomrule
\end{tabular}
}
\label{tab:UCR_size_DM-IM10}
\end{table}

\subsection*{UCR Size-Corrected Power Tables}
\begin{table}[H]
\centering
\caption{Size-Adjusted Power, Unconditional-Rolling DGP, DM-R}
\scalebox{0.7}{
\begin{tabular}{cc|ccccc|ccccc|ccccc}
\toprule
\toprule
&       & \multicolumn{5}{c|}{$h=1$}            & \multicolumn{5}{c|}{$h=3$}            & \multicolumn{5}{c}{$h=12$} \\
\cmidrule{3-17}          &       & \multicolumn{5}{c|}{$P$}              & \multicolumn{5}{c|}{$P$}              & \multicolumn{5}{c}{$P$} \\
$R$   & $\widetilde{R}$ & 25    & 75    & 125   & 175   & 1000  & 25    & 75    & 125   & 175   & 1000  & 25    & 75    & 125   & 175   & 1000 \\
\midrule
25 & 25& \textbf{0.050}& \textbf{0.050}& \textbf{0.050}& \textbf{0.050}& \textbf{0.050}& \textbf{0.050}& \textbf{0.050}& \textbf{0.050}& \textbf{0.050}& \textbf{0.050}& \textbf{0.050}& \textbf{0.050}& \textbf{0.050}& \textbf{0.050}& \textbf{0.050}\\
25 & 75& 0.062& 0.087& 0.077& 0.098& 0.585& 0.078& 0.070& 0.076& 0.087& 0.525& 0.066& 0.065& 0.073& 0.083& 0.416\\
25 & 125& 0.069& 0.114& 0.140& 0.163& 0.818& 0.095& 0.109& 0.123& 0.144& 0.767& 0.069& 0.076& 0.098& 0.107& 0.627\\
25 & 175& 0.074& 0.138& 0.176& 0.216& 0.891& 0.090& 0.126& 0.162& 0.190& 0.855& 0.064& 0.093& 0.121& 0.139& 0.749\\
\midrule
75 & 25& 0.048& 0.087& 0.119& 0.160& 0.613& 0.035& 0.069& 0.101& 0.135& 0.550& 0.046& 0.066& 0.065& 0.092& 0.460\\
\bottomrule
\bottomrule
\end{tabular}
}
\label{tab:UCR_scp_DM-R}
\end{table}

\begin{table}[H]
\centering
\caption{Size-Adjusted Power, Unconditional-Rolling DGP, DM-M}
\scalebox{0.7}{
\begin{tabular}{cc|ccccc|ccccc|ccccc}
\toprule
\toprule
&       & \multicolumn{5}{c|}{$h=1$}            & \multicolumn{5}{c|}{$h=3$}            & \multicolumn{5}{c}{$h=12$} \\
\cmidrule{3-17}          &       & \multicolumn{5}{c|}{$P$}              & \multicolumn{5}{c|}{$P$}              & \multicolumn{5}{c}{$P$} \\
$R$   & $\widetilde{R}$ & 25    & 75    & 125   & 175   & 1000  & 25    & 75    & 125   & 175   & 1000  & 25    & 75    & 125   & 175   & 1000 \\
\midrule
25 & 25& \textbf{0.050}& \textbf{0.050}& \textbf{0.050}& \textbf{0.050}& \textbf{0.050}& \textbf{0.050}& \textbf{0.050}& \textbf{0.050}& \textbf{0.050}& \textbf{0.050}& \textbf{0.050}& \textbf{0.050}& \textbf{0.050}& \textbf{0.050}& \textbf{0.050}\\
25 & 75& 0.062& 0.087& 0.077& 0.098& 0.585& 0.078& 0.070& 0.076& 0.087& 0.525& 0.066& 0.065& 0.073& 0.083& 0.416\\
25 & 125& 0.069& 0.114& 0.140& 0.163& 0.818& 0.095& 0.109& 0.123& 0.144& 0.767& 0.069& 0.076& 0.098& 0.107& 0.627\\
25 & 175& 0.074& 0.138& 0.176& 0.216& 0.891& 0.090& 0.126& 0.162& 0.190& 0.855& 0.064& 0.093& 0.121& 0.139& 0.749\\
\midrule
75 & 25& 0.048& 0.087& 0.119& 0.160& 0.613& 0.035& 0.069& 0.101& 0.135& 0.550& 0.046& 0.066& 0.065& 0.092& 0.460\\
\bottomrule
\bottomrule
\end{tabular}
}
\label{tab:UCR_scp_DM-M}
\end{table}

\begin{table}[H]
\centering
\caption{Size-Adjusted Power, Unconditional-Rolling DGP, DM-NW}
\scalebox{0.7}{
\begin{tabular}{cc|ccccc|ccccc|ccccc}
\toprule
\toprule
&       & \multicolumn{5}{c|}{$h=1$}            & \multicolumn{5}{c|}{$h=3$}            & \multicolumn{5}{c}{$h=12$} \\
\cmidrule{3-17}          &       & \multicolumn{5}{c|}{$P$}              & \multicolumn{5}{c|}{$P$}              & \multicolumn{5}{c}{$P$} \\
$R$   & $\widetilde{R}$ & 25    & 75    & 125   & 175   & 1000  & 25    & 75    & 125   & 175   & 1000  & 25    & 75    & 125   & 175   & 1000 \\
\midrule
25 & 25& \textbf{0.050}& \textbf{0.050}& \textbf{0.050}& \textbf{0.050}& \textbf{0.050}& \textbf{0.050}& \textbf{0.050}& \textbf{0.050}& \textbf{0.050}& \textbf{0.050}& \textbf{0.050}& \textbf{0.050}& \textbf{0.050}& \textbf{0.050}& \textbf{0.050}\\
25 & 75& 0.055& 0.076& 0.074& 0.077& 0.559& 0.078& 0.070& 0.072& 0.086& 0.520& 0.085& 0.067& 0.082& 0.077& 0.417\\
25 & 125& 0.065& 0.101& 0.125& 0.131& 0.800& 0.097& 0.109& 0.122& 0.141& 0.762& 0.093& 0.099& 0.118& 0.120& 0.647\\
25 & 175& 0.069& 0.124& 0.160& 0.180& 0.875& 0.094& 0.127& 0.158& 0.190& 0.853& 0.085& 0.115& 0.156& 0.161& 0.771\\
\midrule
75 & 25& 0.062& 0.104& 0.131& 0.180& 0.629& 0.038& 0.069& 0.103& 0.143& 0.553& 0.041& 0.072& 0.087& 0.132& 0.486\\
\bottomrule
\bottomrule
\end{tabular}
}
\label{tab:UCR_scp_DM-NW}
\end{table}

\begin{table}[H]
\centering
\caption{Size-Adjusted Power, Unconditional-Rolling DGP, DM-NW-L}
\scalebox{0.7}{
\begin{tabular}{cc|ccccc|ccccc|ccccc}
\toprule
\toprule
&       & \multicolumn{5}{c|}{$h=1$}            & \multicolumn{5}{c|}{$h=3$}            & \multicolumn{5}{c}{$h=12$} \\
\cmidrule{3-17}          &       & \multicolumn{5}{c|}{$P$}              & \multicolumn{5}{c|}{$P$}              & \multicolumn{5}{c}{$P$} \\
$R$   & $\widetilde{R}$ & 25    & 75    & 125   & 175   & 1000  & 25    & 75    & 125   & 175   & 1000  & 25    & 75    & 125   & 175   & 1000 \\
\midrule
25 & 25& \textbf{0.050}& \textbf{0.050}& \textbf{0.050}& \textbf{0.050}& \textbf{0.050}& \textbf{0.050}& \textbf{0.050}& \textbf{0.050}& \textbf{0.050}& \textbf{0.050}& \textbf{0.050}& \textbf{0.050}& \textbf{0.050}& \textbf{0.050}& \textbf{0.050}\\
25 & 75& 0.053& 0.052& 0.055& 0.049& 0.438& 0.077& 0.059& 0.058& 0.072& 0.428& 0.088& 0.065& 0.079& 0.083& 0.408\\
25 & 125& 0.067& 0.079& 0.093& 0.096& 0.678& 0.097& 0.096& 0.093& 0.114& 0.654& 0.094& 0.098& 0.122& 0.120& 0.618\\
25 & 175& 0.069& 0.091& 0.124& 0.130& 0.775& 0.098& 0.114& 0.128& 0.150& 0.754& 0.088& 0.114& 0.156& 0.154& 0.737\\
\midrule
75 & 25& 0.065& 0.117& 0.154& 0.210& 0.672& 0.036& 0.092& 0.121& 0.169& 0.592& 0.037& 0.065& 0.079& 0.123& 0.476\\
\bottomrule
\bottomrule
\end{tabular}
}
\label{tab:UCR_scp_DM-NW-L}
\end{table}

\begin{table}[H]
\centering
\caption{Size-Adjusted Power, Unconditional-Rolling DGP, DM-FB}
\scalebox{0.7}{
\begin{tabular}{cc|ccccc|ccccc|ccccc}
\toprule
\toprule
&       & \multicolumn{5}{c|}{$h=1$}            & \multicolumn{5}{c|}{$h=3$}            & \multicolumn{5}{c}{$h=12$} \\
\cmidrule{3-17}          &       & \multicolumn{5}{c|}{$P$}              & \multicolumn{5}{c|}{$P$}              & \multicolumn{5}{c}{$P$} \\
$R$   & $\widetilde{R}$ & 25    & 75    & 125   & 175   & 1000  & 25    & 75    & 125   & 175   & 1000  & 25    & 75    & 125   & 175   & 1000 \\
\midrule
25 & 25& \textbf{0.050}& \textbf{0.050}& \textbf{0.050}& \textbf{0.050}& \textbf{0.050}& \textbf{0.050}& \textbf{0.050}& \textbf{0.050}& \textbf{0.050}& \textbf{0.050}& \textbf{0.050}& \textbf{0.050}& \textbf{0.050}& \textbf{0.050}& \textbf{0.050}\\
25 & 75& 0.053& 0.052& 0.055& 0.049& 0.438& 0.077& 0.059& 0.058& 0.072& 0.428& 0.088& 0.065& 0.079& 0.083& 0.408\\
25 & 125& 0.067& 0.079& 0.093& 0.096& 0.678& 0.097& 0.096& 0.093& 0.114& 0.654& 0.094& 0.098& 0.122& 0.120& 0.618\\
25 & 175& 0.069& 0.091& 0.124& 0.130& 0.775& 0.098& 0.114& 0.128& 0.150& 0.754& 0.088& 0.114& 0.156& 0.154& 0.737\\
\midrule
75 & 25& 0.065& 0.117& 0.154& 0.210& 0.672& 0.036& 0.092& 0.121& 0.169& 0.592& 0.037& 0.065& 0.079& 0.123& 0.476\\
\bottomrule
\bottomrule
\end{tabular}
}
\label{tab:UCR_scp_DM-FB}
\end{table}

\begin{table}[H]
\centering
\caption{Size-Adjusted Power, Unconditional-Rolling DGP, DM-EWC}
\scalebox{0.7}{
\begin{tabular}{cc|ccccc|ccccc|ccccc}
\toprule
\toprule
&       & \multicolumn{5}{c|}{$h=1$}            & \multicolumn{5}{c|}{$h=3$}            & \multicolumn{5}{c}{$h=12$} \\
\cmidrule{3-17}          &       & \multicolumn{5}{c|}{$P$}              & \multicolumn{5}{c|}{$P$}              & \multicolumn{5}{c}{$P$} \\
$R$   & $\widetilde{R}$ & 25    & 75    & 125   & 175   & 1000  & 25    & 75    & 125   & 175   & 1000  & 25    & 75    & 125   & 175   & 1000 \\
\midrule
25 & 25& \textbf{0.050}& \textbf{0.050}& \textbf{0.050}& \textbf{0.050}& \textbf{0.050}& \textbf{0.050}& \textbf{0.050}& \textbf{0.050}& \textbf{0.050}& \textbf{0.050}& \textbf{0.050}& \textbf{0.050}& \textbf{0.050}& \textbf{0.050}& \textbf{0.050}\\
25 & 75& 0.051& 0.046& 0.049& 0.047& 0.432& 0.068& 0.059& 0.062& 0.071& 0.425& 0.079& 0.064& 0.083& 0.078& 0.384\\
25 & 125& 0.060& 0.067& 0.082& 0.087& 0.676& 0.080& 0.099& 0.097& 0.109& 0.659& 0.089& 0.091& 0.120& 0.112& 0.596\\
25 & 175& 0.060& 0.076& 0.110& 0.120& 0.772& 0.086& 0.110& 0.128& 0.147& 0.753& 0.083& 0.107& 0.157& 0.146& 0.725\\
\midrule
75 & 25& 0.061& 0.112& 0.144& 0.205& 0.683& 0.034& 0.091& 0.109& 0.151& 0.596& 0.038& 0.061& 0.067& 0.116& 0.488\\
\bottomrule
\bottomrule
\end{tabular}
}
\label{tab:UCR_scp_DM-EWC}
\end{table}

\begin{table}[H]
\centering
\caption{Size-Adjusted Power, Unconditional-Rolling DGP, DM-IM2}
\scalebox{0.7}{
\begin{tabular}{cc|ccccc|ccccc|ccccc}
\toprule
\toprule
&       & \multicolumn{5}{c|}{$h=1$}            & \multicolumn{5}{c|}{$h=3$}            & \multicolumn{5}{c}{$h=12$} \\
\cmidrule{3-17}          &       & \multicolumn{5}{c|}{$P$}              & \multicolumn{5}{c|}{$P$}              & \multicolumn{5}{c}{$P$} \\
$R$   & $\widetilde{R}$ & 25    & 75    & 125   & 175   & 1000  & 25    & 75    & 125   & 175   & 1000  & 25    & 75    & 125   & 175   & 1000 \\
\midrule
25 & 25& \textbf{0.050}& \textbf{0.050}& \textbf{0.050}& \textbf{0.050}& \textbf{0.050}& \textbf{0.050}& \textbf{0.050}& \textbf{0.050}& \textbf{0.050}& \textbf{0.050}& \textbf{0.050}& \textbf{0.050}& \textbf{0.050}& \textbf{0.050}& \textbf{0.050}\\
25 & 75& 0.057& 0.042& 0.044& 0.055& 0.125& 0.049& 0.050& 0.047& 0.058& 0.110& 0.062& 0.054& 0.062& 0.051& 0.123\\
25 & 125& 0.063& 0.040& 0.049& 0.069& 0.142& 0.053& 0.048& 0.054& 0.063& 0.143& 0.062& 0.060& 0.066& 0.057& 0.152\\
25 & 175& 0.063& 0.051& 0.059& 0.068& 0.158& 0.056& 0.058& 0.060& 0.066& 0.148& 0.063& 0.071& 0.074& 0.062& 0.161\\
\midrule
75 & 25& 0.051& 0.085& 0.087& 0.087& 0.172& 0.049& 0.071& 0.076& 0.070& 0.147& 0.054& 0.051& 0.060& 0.061& 0.115\\
\bottomrule
\bottomrule
\end{tabular}
}
\label{tab:UCR_scp_DM-IM2}
\end{table}

\begin{table}[H]
\centering
\caption{Size-Adjusted Power, Unconditional-Rolling DGP, DM-IM5}
\scalebox{0.7}{
\begin{tabular}{cc|ccccc|ccccc|ccccc}
\toprule
\toprule
&       & \multicolumn{5}{c|}{$h=1$}            & \multicolumn{5}{c|}{$h=3$}            & \multicolumn{5}{c}{$h=12$} \\
\cmidrule{3-17}          &       & \multicolumn{5}{c|}{$P$}              & \multicolumn{5}{c|}{$P$}              & \multicolumn{5}{c}{$P$} \\
$R$   & $\widetilde{R}$ & 25    & 75    & 125   & 175   & 1000  & 25    & 75    & 125   & 175   & 1000  & 25    & 75    & 125   & 175   & 1000 \\
\midrule
25 & 25& \textbf{0.050}& \textbf{0.050}& \textbf{0.050}& \textbf{0.050}& \textbf{0.050}& \textbf{0.050}& \textbf{0.050}& \textbf{0.050}& \textbf{0.050}& \textbf{0.050}& \textbf{0.050}& \textbf{0.050}& \textbf{0.050}& \textbf{0.050}& \textbf{0.050}\\
25 & 75& 0.047& 0.047& 0.034& 0.043& 0.250& 0.067& 0.050& 0.052& 0.049& 0.260& 0.084& 0.063& 0.075& 0.080& 0.273\\
25 & 125& 0.055& 0.050& 0.054& 0.056& 0.402& 0.079& 0.069& 0.066& 0.074& 0.415& 0.094& 0.091& 0.109& 0.108& 0.422\\
25 & 175& 0.062& 0.062& 0.074& 0.073& 0.487& 0.090& 0.083& 0.089& 0.089& 0.477& 0.083& 0.098& 0.132& 0.129& 0.484\\
\midrule
75 & 25& 0.065& 0.108& 0.152& 0.210& 0.439& 0.039& 0.097& 0.117& 0.148& 0.387& 0.032& 0.060& 0.073& 0.094& 0.267\\
\bottomrule
\bottomrule
\end{tabular}
}
\label{tab:UCR_scp_DM-IM5}
\end{table}

\begin{table}[H]
\centering
\caption{Size-Adjusted Power, Unconditional-Rolling DGP, DM-IM10}
\scalebox{0.7}{
\begin{tabular}{cc|ccccc|ccccc|ccccc}
\toprule
\toprule
&       & \multicolumn{5}{c|}{$h=1$}            & \multicolumn{5}{c|}{$h=3$}            & \multicolumn{5}{c}{$h=12$} \\
\cmidrule{3-17}          &       & \multicolumn{5}{c|}{$P$}              & \multicolumn{5}{c|}{$P$}              & \multicolumn{5}{c}{$P$} \\
$R$   & $\widetilde{R}$ & 25    & 75    & 125   & 175   & 1000  & 25    & 75    & 125   & 175   & 1000  & 25    & 75    & 125   & 175   & 1000 \\
\midrule
25 & 25& \textbf{0.050}& \textbf{0.050}& \textbf{0.050}& \textbf{0.050}& \textbf{0.050}& \textbf{0.050}& \textbf{0.050}& \textbf{0.050}& \textbf{0.050}& \textbf{0.050}& \textbf{0.050}& \textbf{0.050}& \textbf{0.050}& \textbf{0.050}& \textbf{0.050}\\
25 & 75& 0.054& 0.062& 0.046& 0.047& 0.365& 0.072& 0.062& 0.060& 0.063& 0.340& 0.079& 0.069& 0.077& 0.087& 0.351\\
25 & 125& 0.059& 0.075& 0.075& 0.081& 0.569& 0.092& 0.094& 0.089& 0.101& 0.538& 0.085& 0.095& 0.114& 0.112& 0.533\\
25 & 175& 0.060& 0.087& 0.099& 0.104& 0.657& 0.090& 0.110& 0.116& 0.125& 0.621& 0.078& 0.109& 0.150& 0.144& 0.633\\
\midrule
75 & 25& 0.057& 0.111& 0.152& 0.197& 0.547& 0.036& 0.085& 0.116& 0.153& 0.488& 0.042& 0.071& 0.080& 0.101& 0.355\\
\bottomrule
\bottomrule
\end{tabular}
}
\label{tab:UCR_scp_DM-IM10}
\end{table}

\subsection*{CR Size Tables}
\begin{table}[H]
\centering
\caption{Size, Conditional-Rolling DGP, DM-R}
\scalebox{0.7}{
\begin{tabular}{cc|ccccc|ccccc|ccccc}
\toprule
\toprule
&       & \multicolumn{5}{c|}{$h=1$}            & \multicolumn{5}{c|}{$h=3$}            & \multicolumn{5}{c}{$h=12$} \\
\cmidrule{3-17}          &       & \multicolumn{5}{c|}{$P$}              & \multicolumn{5}{c|}{$P$}              & \multicolumn{5}{c}{$P$} \\
$R$   & $\widetilde{R}$ & 25    & 75    & 125   & 175   & 1000  & 25    & 75    & 125   & 175   & 1000  & 25    & 75    & 125   & 175   & 1000 \\
\midrule
25 & 25& \textbf{0.061}& \textbf{0.054}& \textbf{0.046}& \textbf{0.051}& \textbf{0.051}& \textbf{0.111}& \textbf{0.061}& \textbf{0.054}& \textbf{0.054}& \textbf{0.053}& \textbf{0.339}& \textbf{0.148}& \textbf{0.111}& \textbf{0.099}& \textbf{0.057}\\
25 & 75& 0.111& 0.097& 0.086& 0.081& 0.075& 0.168& 0.105& 0.084& 0.075& 0.058& 0.366& 0.190& 0.137& 0.113& 0.059\\
25 & 125& 0.132& 0.159& 0.129& 0.115& 0.107& 0.213& 0.159& 0.129& 0.113& 0.093& 0.382& 0.238& 0.159& 0.146& 0.079\\
25 & 175& 0.153& 0.182& 0.167& 0.143& 0.119& 0.218& 0.200& 0.167& 0.139& 0.108& 0.401& 0.281& 0.220& 0.157& 0.097\\
\midrule
75 & 25& 0.070& 0.092& 0.123& 0.144& 0.503& 0.114& 0.103& 0.124& 0.151& 0.481& 0.331& 0.201& 0.184& 0.181& 0.427\\
\bottomrule
\bottomrule
\end{tabular}
}
\label{tab:CR_size_DM-R}
\end{table}

\begin{table}[H]
\centering
\caption{Size, Conditional-Rolling DGP, DM-M}
\scalebox{0.7}{
\begin{tabular}{cc|ccccc|ccccc|ccccc}
\toprule
\toprule
&       & \multicolumn{5}{c|}{$h=1$}            & \multicolumn{5}{c|}{$h=3$}            & \multicolumn{5}{c}{$h=12$} \\
\cmidrule{3-17}          &       & \multicolumn{5}{c|}{$P$}              & \multicolumn{5}{c|}{$P$}              & \multicolumn{5}{c}{$P$} \\
$R$   & $\widetilde{R}$ & 25    & 75    & 125   & 175   & 1000  & 25    & 75    & 125   & 175   & 1000  & 25    & 75    & 125   & 175   & 1000 \\
\midrule
25 & 25& \textbf{0.044}& \textbf{0.048}& \textbf{0.042}& \textbf{0.050}& \textbf{0.051}& \textbf{0.063}& \textbf{0.049}& \textbf{0.047}& \textbf{0.049}& \textbf{0.053}& \textbf{0.117}& \textbf{0.097}& \textbf{0.083}& \textbf{0.078}& \textbf{0.055}\\
25 & 75& 0.087& 0.089& 0.083& 0.077& 0.075& 0.109& 0.087& 0.076& 0.068& 0.057& 0.137& 0.128& 0.105& 0.085& 0.055\\
25 & 125& 0.106& 0.149& 0.124& 0.112& 0.106& 0.145& 0.137& 0.117& 0.105& 0.091& 0.153& 0.158& 0.121& 0.117& 0.076\\
25 & 175& 0.125& 0.171& 0.161& 0.140& 0.118& 0.156& 0.174& 0.154& 0.129& 0.107& 0.167& 0.203& 0.173& 0.129& 0.094\\
\midrule
75 & 25& 0.049& 0.084& 0.115& 0.140& 0.502& 0.066& 0.084& 0.111& 0.140& 0.479& 0.117& 0.135& 0.140& 0.146& 0.419\\
\bottomrule
\bottomrule
\end{tabular}
}
\label{tab:CR_size_DM-M}
\end{table}

\begin{table}[H]
\centering
\caption{Size, Conditional-Rolling DGP, DM-NW}
\scalebox{0.7}{
\begin{tabular}{cc|ccccc|ccccc|ccccc}
\toprule
\toprule
&       & \multicolumn{5}{c|}{$h=1$}            & \multicolumn{5}{c|}{$h=3$}            & \multicolumn{5}{c}{$h=12$} \\
\cmidrule{3-17}          &       & \multicolumn{5}{c|}{$P$}              & \multicolumn{5}{c|}{$P$}              & \multicolumn{5}{c}{$P$} \\
$R$   & $\widetilde{R}$ & 25    & 75    & 125   & 175   & 1000  & 25    & 75    & 125   & 175   & 1000  & 25    & 75    & 125   & 175   & 1000 \\
\midrule
25 & 25& \textbf{0.092}& \textbf{0.073}& \textbf{0.064}& \textbf{0.066}& \textbf{0.056}& \textbf{0.131}& \textbf{0.087}& \textbf{0.079}& \textbf{0.079}& \textbf{0.069}& \textbf{0.170}& \textbf{0.101}& \textbf{0.090}& \textbf{0.086}& \textbf{0.073}\\
25 & 75& 0.134& 0.109& 0.093& 0.081& 0.064& 0.198& 0.141& 0.111& 0.104& 0.071& 0.220& 0.145& 0.119& 0.110& 0.090\\
25 & 125& 0.160& 0.163& 0.128& 0.115& 0.088& 0.243& 0.206& 0.162& 0.144& 0.103& 0.233& 0.207& 0.164& 0.149& 0.120\\
25 & 175& 0.183& 0.188& 0.165& 0.137& 0.099& 0.251& 0.246& 0.201& 0.174& 0.122& 0.249& 0.247& 0.220& 0.181& 0.149\\
\midrule
75 & 25& 0.103& 0.130& 0.161& 0.185& 0.538& 0.136& 0.143& 0.171& 0.197& 0.539& 0.180& 0.158& 0.176& 0.204& 0.496\\
\bottomrule
\bottomrule
\end{tabular}
}
\label{tab:CR_size_DM-NW}
\end{table}

\begin{table}[H]
\centering
\caption{Size, Conditional-Rolling DGP, DM-NW-L}
\scalebox{0.7}{
\begin{tabular}{cc|ccccc|ccccc|ccccc}
\toprule
\toprule
&       & \multicolumn{5}{c|}{$h=1$}            & \multicolumn{5}{c|}{$h=3$}            & \multicolumn{5}{c}{$h=12$} \\
\cmidrule{3-17}          &       & \multicolumn{5}{c|}{$P$}              & \multicolumn{5}{c|}{$P$}              & \multicolumn{5}{c}{$P$} \\
$R$   & $\widetilde{R}$ & 25    & 75    & 125   & 175   & 1000  & 25    & 75    & 125   & 175   & 1000  & 25    & 75    & 125   & 175   & 1000 \\
\midrule
25 & 25& \textbf{0.149}& \textbf{0.120}& \textbf{0.110}& \textbf{0.112}& \textbf{0.089}& \textbf{0.160}& \textbf{0.119}& \textbf{0.110}& \textbf{0.107}& \textbf{0.089}& \textbf{0.177}& \textbf{0.098}& \textbf{0.087}& \textbf{0.079}& \textbf{0.066}\\
25 & 75& 0.196& 0.137& 0.110& 0.091& 0.052& 0.226& 0.146& 0.118& 0.098& 0.048& 0.227& 0.137& 0.103& 0.095& 0.056\\
25 & 125& 0.221& 0.189& 0.133& 0.114& 0.049& 0.270& 0.203& 0.147& 0.126& 0.049& 0.239& 0.191& 0.137& 0.130& 0.059\\
25 & 175& 0.240& 0.211& 0.162& 0.133& 0.045& 0.275& 0.236& 0.183& 0.140& 0.050& 0.258& 0.236& 0.197& 0.146& 0.072\\
\midrule
75 & 25& 0.166& 0.213& 0.248& 0.299& 0.646& 0.169& 0.194& 0.223& 0.273& 0.617& 0.176& 0.141& 0.156& 0.176& 0.483\\
\bottomrule
\bottomrule
\end{tabular}
}
\label{tab:CR_size_DM-NW-L}
\end{table}

\begin{table}[H]
\centering
\caption{Size, Conditional-Rolling DGP, DM-FB}
\scalebox{0.7}{
\begin{tabular}{cc|ccccc|ccccc|ccccc}
\toprule
\toprule
&       & \multicolumn{5}{c|}{$h=1$}            & \multicolumn{5}{c|}{$h=3$}            & \multicolumn{5}{c}{$h=12$} \\
\cmidrule{3-17}          &       & \multicolumn{5}{c|}{$P$}              & \multicolumn{5}{c|}{$P$}              & \multicolumn{5}{c}{$P$} \\
$R$   & $\widetilde{R}$ & 25    & 75    & 125   & 175   & 1000  & 25    & 75    & 125   & 175   & 1000  & 25    & 75    & 125   & 175   & 1000 \\
\midrule
25 & 25& \textbf{0.042}& \textbf{0.061}& \textbf{0.065}& \textbf{0.073}& \textbf{0.075}& \textbf{0.045}& \textbf{0.051}& \textbf{0.057}& \textbf{0.063}& \textbf{0.075}& \textbf{0.064}& \textbf{0.043}& \textbf{0.044}& \textbf{0.039}& \textbf{0.053}\\
25 & 75& 0.067& 0.063& 0.056& 0.054& 0.042& 0.089& 0.069& 0.062& 0.055& 0.037& 0.095& 0.066& 0.054& 0.052& 0.040\\
25 & 125& 0.084& 0.095& 0.077& 0.068& 0.038& 0.123& 0.103& 0.082& 0.074& 0.034& 0.106& 0.095& 0.076& 0.074& 0.047\\
25 & 175& 0.102& 0.111& 0.091& 0.079& 0.035& 0.130& 0.136& 0.101& 0.083& 0.038& 0.118& 0.138& 0.121& 0.089& 0.054\\
\midrule
75 & 25& 0.053& 0.114& 0.174& 0.218& 0.615& 0.050& 0.096& 0.145& 0.196& 0.584& 0.061& 0.068& 0.087& 0.112& 0.440\\
\bottomrule
\bottomrule
\end{tabular}
}
\label{tab:CR_size_DM-FB}
\end{table}

\begin{table}[H]
\centering
\caption{Size, Conditional-Rolling DGP, DM-EWC}
\scalebox{0.7}{
\begin{tabular}{cc|ccccc|ccccc|ccccc}
\toprule
\toprule
&       & \multicolumn{5}{c|}{$h=1$}            & \multicolumn{5}{c|}{$h=3$}            & \multicolumn{5}{c}{$h=12$} \\
\cmidrule{3-17}          &       & \multicolumn{5}{c|}{$P$}              & \multicolumn{5}{c|}{$P$}              & \multicolumn{5}{c}{$P$} \\
$R$   & $\widetilde{R}$ & 25    & 75    & 125   & 175   & 1000  & 25    & 75    & 125   & 175   & 1000  & 25    & 75    & 125   & 175   & 1000 \\
\midrule
25 & 25& \textbf{0.053}& \textbf{0.066}& \textbf{0.071}& \textbf{0.078}& \textbf{0.071}& \textbf{0.036}& \textbf{0.047}& \textbf{0.055}& \textbf{0.060}& \textbf{0.073}& \textbf{0.041}& \textbf{0.039}& \textbf{0.041}& \textbf{0.038}& \textbf{0.051}\\
25 & 75& 0.063& 0.066& 0.062& 0.059& 0.034& 0.061& 0.064& 0.063& 0.055& 0.033& 0.056& 0.057& 0.049& 0.053& 0.045\\
25 & 125& 0.070& 0.100& 0.086& 0.078& 0.041& 0.081& 0.095& 0.081& 0.080& 0.039& 0.065& 0.088& 0.075& 0.070& 0.060\\
25 & 175& 0.080& 0.111& 0.098& 0.090& 0.041& 0.080& 0.123& 0.105& 0.090& 0.047& 0.071& 0.122& 0.115& 0.092& 0.076\\
\midrule
75 & 25& 0.053& 0.123& 0.172& 0.227& 0.616& 0.037& 0.086& 0.129& 0.179& 0.575& 0.036& 0.052& 0.071& 0.095& 0.423\\
\bottomrule
\bottomrule
\end{tabular}
}
\label{tab:CR_size_DM-EWC}
\end{table}

\begin{table}[H]
\centering
\caption{Size, Conditional-Rolling DGP, DM-IM2}
\scalebox{0.7}{
\begin{tabular}{cc|ccccc|ccccc|ccccc}
\toprule
\toprule
&       & \multicolumn{5}{c|}{$h=1$}            & \multicolumn{5}{c|}{$h=3$}            & \multicolumn{5}{c}{$h=12$} \\
\cmidrule{3-17}          &       & \multicolumn{5}{c|}{$P$}              & \multicolumn{5}{c|}{$P$}              & \multicolumn{5}{c}{$P$} \\
$R$   & $\widetilde{R}$ & 25    & 75    & 125   & 175   & 1000  & 25    & 75    & 125   & 175   & 1000  & 25    & 75    & 125   & 175   & 1000 \\
\midrule
25 & 25& \textbf{0.045}& \textbf{0.055}& \textbf{0.056}& \textbf{0.057}& \textbf{0.054}& \textbf{0.041}& \textbf{0.050}& \textbf{0.054}& \textbf{0.053}& \textbf{0.057}& \textbf{0.047}& \textbf{0.040}& \textbf{0.045}& \textbf{0.041}& \textbf{0.042}\\
25 & 75& 0.055& 0.048& 0.045& 0.056& 0.054& 0.051& 0.048& 0.045& 0.049& 0.050& 0.048& 0.045& 0.047& 0.044& 0.049\\
25 & 125& 0.055& 0.045& 0.037& 0.044& 0.054& 0.061& 0.050& 0.036& 0.043& 0.057& 0.054& 0.049& 0.042& 0.041& 0.057\\
25 & 175& 0.058& 0.048& 0.045& 0.042& 0.055& 0.065& 0.052& 0.041& 0.037& 0.057& 0.054& 0.048& 0.043& 0.037& 0.053\\
\midrule
75 & 25& 0.054& 0.079& 0.099& 0.093& 0.160& 0.048& 0.069& 0.075& 0.087& 0.154& 0.048& 0.052& 0.057& 0.063& 0.113\\
\bottomrule
\bottomrule
\end{tabular}
}
\label{tab:CR_size_DM-IM2}
\end{table}

\begin{table}[H]
\centering
\caption{Size, Conditional-Rolling DGP, DM-IM5}
\scalebox{0.7}{
\begin{tabular}{cc|ccccc|ccccc|ccccc}
\toprule
\toprule
&       & \multicolumn{5}{c|}{$h=1$}            & \multicolumn{5}{c|}{$h=3$}            & \multicolumn{5}{c}{$h=12$} \\
\cmidrule{3-17}          &       & \multicolumn{5}{c|}{$P$}              & \multicolumn{5}{c|}{$P$}              & \multicolumn{5}{c}{$P$} \\
$R$   & $\widetilde{R}$ & 25    & 75    & 125   & 175   & 1000  & 25    & 75    & 125   & 175   & 1000  & 25    & 75    & 125   & 175   & 1000 \\
\midrule
25 & 25& \textbf{0.049}& \textbf{0.078}& \textbf{0.089}& \textbf{0.104}& \textbf{0.073}& \textbf{0.040}& \textbf{0.054}& \textbf{0.066}& \textbf{0.077}& \textbf{0.079}& \textbf{0.056}& \textbf{0.039}& \textbf{0.040}& \textbf{0.039}& \textbf{0.055}\\
25 & 75& 0.070& 0.063& 0.048& 0.053& 0.084& 0.076& 0.055& 0.052& 0.048& 0.079& 0.077& 0.053& 0.043& 0.044& 0.062\\
25 & 125& 0.077& 0.076& 0.054& 0.046& 0.080& 0.097& 0.077& 0.045& 0.047& 0.077& 0.082& 0.071& 0.047& 0.038& 0.065\\
25 & 175& 0.091& 0.086& 0.056& 0.048& 0.068& 0.105& 0.100& 0.053& 0.043& 0.065& 0.097& 0.095& 0.068& 0.047& 0.070\\
\midrule
75 & 25& 0.064& 0.130& 0.188& 0.250& 0.492& 0.047& 0.095& 0.144& 0.196& 0.455& 0.050& 0.052& 0.067& 0.077& 0.298\\
\bottomrule
\bottomrule
\end{tabular}
}
\label{tab:CR_size_DM-IM5}
\end{table}

\begin{table}[H]
\centering
\caption{Size, Conditional-Rolling DGP, DM-IM10}
\scalebox{0.7}{
\begin{tabular}{cc|ccccc|ccccc|ccccc}
\toprule
\toprule
&       & \multicolumn{5}{c|}{$h=1$}            & \multicolumn{5}{c|}{$h=3$}            & \multicolumn{5}{c}{$h=12$} \\
\cmidrule{3-17}          &       & \multicolumn{5}{c|}{$P$}              & \multicolumn{5}{c|}{$P$}              & \multicolumn{5}{c}{$P$} \\
$R$   & $\widetilde{R}$ & 25    & 75    & 125   & 175   & 1000  & 25    & 75    & 125   & 175   & 1000  & 25    & 75    & 125   & 175   & 1000 \\
\midrule
25 & 25& \textbf{0.048}& \textbf{0.068}& \textbf{0.077}& \textbf{0.086}& \textbf{0.089}& \textbf{0.070}& \textbf{0.058}& \textbf{0.069}& \textbf{0.073}& \textbf{0.079}& \textbf{0.109}& \textbf{0.053}& \textbf{0.048}& \textbf{0.047}& \textbf{0.060}\\
25 & 75& 0.078& 0.083& 0.067& 0.062& 0.080& 0.129& 0.085& 0.064& 0.059& 0.077& 0.147& 0.080& 0.057& 0.056& 0.063\\
25 & 125& 0.091& 0.110& 0.085& 0.074& 0.065& 0.166& 0.121& 0.090& 0.072& 0.062& 0.161& 0.112& 0.079& 0.073& 0.054\\
25 & 175& 0.113& 0.126& 0.098& 0.078& 0.044& 0.172& 0.149& 0.108& 0.082& 0.046& 0.178& 0.148& 0.120& 0.086& 0.049\\
\midrule
75 & 25& 0.060& 0.126& 0.178& 0.235& 0.579& 0.074& 0.099& 0.144& 0.202& 0.548& 0.116& 0.077& 0.096& 0.109& 0.394\\
\bottomrule
\bottomrule
\end{tabular}
}
\label{tab:CR_size_DM-IM10}
\end{table}

\subsection*{CR Size-Corrected Power Tables}
\begin{table}[H]
\centering
\caption{Size-Adjusted Power, Conditional-Rolling DGP, DM-R}
\scalebox{0.7}{
\begin{tabular}{cc|ccccc|ccccc|ccccc}
\toprule
\toprule
&       & \multicolumn{5}{c|}{$h=1$}            & \multicolumn{5}{c|}{$h=3$}            & \multicolumn{5}{c}{$h=12$} \\
\cmidrule{3-17}          &       & \multicolumn{5}{c|}{$P$}              & \multicolumn{5}{c|}{$P$}              & \multicolumn{5}{c}{$P$} \\
$R$   & $\widetilde{R}$ & 25    & 75    & 125   & 175   & 1000  & 25    & 75    & 125   & 175   & 1000  & 25    & 75    & 125   & 175   & 1000 \\
\midrule
25 & 25& \textbf{0.050}& \textbf{0.050}& \textbf{0.050}& \textbf{0.050}& \textbf{0.050}& \textbf{0.050}& \textbf{0.050}& \textbf{0.050}& \textbf{0.050}& \textbf{0.050}& \textbf{0.050}& \textbf{0.050}& \textbf{0.050}& \textbf{0.050}& \textbf{0.050}\\
25 & 75& 0.095& 0.091& 0.091& 0.077& 0.073& 0.091& 0.088& 0.080& 0.072& 0.054& 0.064& 0.058& 0.062& 0.057& 0.049\\
25 & 125& 0.115& 0.153& 0.134& 0.112& 0.105& 0.119& 0.138& 0.123& 0.109& 0.087& 0.073& 0.081& 0.074& 0.081& 0.070\\
25 & 175& 0.133& 0.175& 0.174& 0.140& 0.115& 0.130& 0.175& 0.160& 0.135& 0.102& 0.086& 0.108& 0.106& 0.090& 0.088\\
\midrule
75 & 25& 0.042& 0.087& 0.115& 0.149& 0.509& 0.046& 0.072& 0.113& 0.132& 0.473& 0.045& 0.062& 0.079& 0.107& 0.410\\
\bottomrule
\bottomrule
\end{tabular}
}
\label{tab:CR_scp_DM-R}
\end{table}

\begin{table}[H]
\centering
\caption{Size-Adjusted Power, Conditional-Rolling DGP, DM-M}
\scalebox{0.7}{
\begin{tabular}{cc|ccccc|ccccc|ccccc}
\toprule
\toprule
&       & \multicolumn{5}{c|}{$h=1$}            & \multicolumn{5}{c|}{$h=3$}            & \multicolumn{5}{c}{$h=12$} \\
\cmidrule{3-17}          &       & \multicolumn{5}{c|}{$P$}              & \multicolumn{5}{c|}{$P$}              & \multicolumn{5}{c}{$P$} \\
$R$   & $\widetilde{R}$ & 25    & 75    & 125   & 175   & 1000  & 25    & 75    & 125   & 175   & 1000  & 25    & 75    & 125   & 175   & 1000 \\
\midrule
25 & 25& \textbf{0.050}& \textbf{0.050}& \textbf{0.050}& \textbf{0.050}& \textbf{0.050}& \textbf{0.050}& \textbf{0.050}& \textbf{0.050}& \textbf{0.050}& \textbf{0.050}& \textbf{0.050}& \textbf{0.050}& \textbf{0.050}& \textbf{0.050}& \textbf{0.050}\\
25 & 75& 0.095& 0.091& 0.091& 0.077& 0.073& 0.091& 0.088& 0.080& 0.072& 0.054& 0.064& 0.058& 0.062& 0.057& 0.049\\
25 & 125& 0.115& 0.153& 0.134& 0.112& 0.105& 0.119& 0.138& 0.123& 0.109& 0.087& 0.073& 0.081& 0.074& 0.081& 0.070\\
25 & 175& 0.133& 0.175& 0.174& 0.140& 0.115& 0.130& 0.175& 0.160& 0.135& 0.102& 0.086& 0.108& 0.106& 0.090& 0.088\\
\midrule
75 & 25& 0.042& 0.087& 0.115& 0.149& 0.509& 0.046& 0.072& 0.113& 0.132& 0.473& 0.045& 0.062& 0.079& 0.107& 0.410\\
\bottomrule
\bottomrule
\end{tabular}
}
\label{tab:CR_scp_DM-M}
\end{table}

\begin{table}[H]
\centering
\caption{Size-Adjusted Power, Conditional-Rolling DGP, DM-NW}
\scalebox{0.7}{
\begin{tabular}{cc|ccccc|ccccc|ccccc}
\toprule
\toprule
&       & \multicolumn{5}{c|}{$h=1$}            & \multicolumn{5}{c|}{$h=3$}            & \multicolumn{5}{c}{$h=12$} \\
\cmidrule{3-17}          &       & \multicolumn{5}{c|}{$P$}              & \multicolumn{5}{c|}{$P$}              & \multicolumn{5}{c}{$P$} \\
$R$   & $\widetilde{R}$ & 25    & 75    & 125   & 175   & 1000  & 25    & 75    & 125   & 175   & 1000  & 25    & 75    & 125   & 175   & 1000 \\
\midrule
25 & 25& \textbf{0.050}& \textbf{0.050}& \textbf{0.050}& \textbf{0.050}& \textbf{0.050}& \textbf{0.050}& \textbf{0.050}& \textbf{0.050}& \textbf{0.050}& \textbf{0.050}& \textbf{0.050}& \textbf{0.050}& \textbf{0.050}& \textbf{0.050}& \textbf{0.050}\\
25 & 75& 0.089& 0.076& 0.073& 0.059& 0.057& 0.103& 0.085& 0.078& 0.068& 0.048& 0.079& 0.082& 0.071& 0.069& 0.061\\
25 & 125& 0.104& 0.117& 0.105& 0.090& 0.080& 0.138& 0.140& 0.117& 0.106& 0.074& 0.089& 0.122& 0.105& 0.098& 0.088\\
25 & 175& 0.129& 0.144& 0.138& 0.103& 0.089& 0.149& 0.176& 0.153& 0.130& 0.089& 0.106& 0.164& 0.157& 0.127& 0.112\\
\midrule
75 & 25& 0.053& 0.095& 0.129& 0.162& 0.522& 0.043& 0.074& 0.114& 0.138& 0.480& 0.044& 0.066& 0.108& 0.140& 0.421\\
\bottomrule
\bottomrule
\end{tabular}
}
\label{tab:CR_scp_DM-NW}
\end{table}

\begin{table}[H]
\centering
\caption{Size-Adjusted Power, Conditional-Rolling DGP, DM-NW-L}
\scalebox{0.7}{
\begin{tabular}{cc|ccccc|ccccc|ccccc}
\toprule
\toprule
&       & \multicolumn{5}{c|}{$h=1$}            & \multicolumn{5}{c|}{$h=3$}            & \multicolumn{5}{c}{$h=12$} \\
\cmidrule{3-17}          &       & \multicolumn{5}{c|}{$P$}              & \multicolumn{5}{c|}{$P$}              & \multicolumn{5}{c}{$P$} \\
$R$   & $\widetilde{R}$ & 25    & 75    & 125   & 175   & 1000  & 25    & 75    & 125   & 175   & 1000  & 25    & 75    & 125   & 175   & 1000 \\
\midrule
25 & 25& \textbf{0.050}& \textbf{0.050}& \textbf{0.050}& \textbf{0.050}& \textbf{0.050}& \textbf{0.050}& \textbf{0.050}& \textbf{0.050}& \textbf{0.050}& \textbf{0.050}& \textbf{0.050}& \textbf{0.050}& \textbf{0.050}& \textbf{0.050}& \textbf{0.050}\\
25 & 75& 0.077& 0.050& 0.040& 0.033& 0.025& 0.095& 0.067& 0.054& 0.041& 0.021& 0.080& 0.073& 0.064& 0.063& 0.038\\
25 & 125& 0.093& 0.074& 0.060& 0.039& 0.021& 0.130& 0.102& 0.071& 0.058& 0.018& 0.087& 0.105& 0.089& 0.091& 0.044\\
25 & 175& 0.112& 0.089& 0.070& 0.046& 0.018& 0.137& 0.134& 0.092& 0.064& 0.016& 0.103& 0.146& 0.138& 0.105& 0.052\\
\midrule
75 & 25& 0.062& 0.110& 0.154& 0.192& 0.572& 0.047& 0.097& 0.135& 0.163& 0.536& 0.040& 0.068& 0.097& 0.137& 0.424\\
\bottomrule
\bottomrule
\end{tabular}
}
\label{tab:CR_scp_DM-NW-L}
\end{table}

\begin{table}[H]
\centering
\caption{Size-Adjusted Power, Conditional-Rolling DGP, DM-FB}
\scalebox{0.7}{
\begin{tabular}{cc|ccccc|ccccc|ccccc}
\toprule
\toprule
&       & \multicolumn{5}{c|}{$h=1$}            & \multicolumn{5}{c|}{$h=3$}            & \multicolumn{5}{c}{$h=12$} \\
\cmidrule{3-17}          &       & \multicolumn{5}{c|}{$P$}              & \multicolumn{5}{c|}{$P$}              & \multicolumn{5}{c}{$P$} \\
$R$   & $\widetilde{R}$ & 25    & 75    & 125   & 175   & 1000  & 25    & 75    & 125   & 175   & 1000  & 25    & 75    & 125   & 175   & 1000 \\
\midrule
25 & 25& \textbf{0.050}& \textbf{0.050}& \textbf{0.050}& \textbf{0.050}& \textbf{0.050}& \textbf{0.050}& \textbf{0.050}& \textbf{0.050}& \textbf{0.050}& \textbf{0.050}& \textbf{0.050}& \textbf{0.050}& \textbf{0.050}& \textbf{0.050}& \textbf{0.050}\\
25 & 75& 0.077& 0.050& 0.040& 0.033& 0.025& 0.095& 0.067& 0.054& 0.041& 0.021& 0.080& 0.073& 0.064& 0.063& 0.038\\
25 & 125& 0.093& 0.074& 0.060& 0.039& 0.021& 0.130& 0.102& 0.071& 0.058& 0.018& 0.087& 0.105& 0.089& 0.091& 0.044\\
25 & 175& 0.112& 0.089& 0.070& 0.046& 0.018& 0.137& 0.134& 0.092& 0.064& 0.016& 0.103& 0.146& 0.138& 0.105& 0.052\\
\midrule
75 & 25& 0.062& 0.110& 0.154& 0.192& 0.572& 0.047& 0.097& 0.135& 0.163& 0.536& 0.040& 0.068& 0.097& 0.137& 0.424\\
\bottomrule
\bottomrule
\end{tabular}
}
\label{tab:CR_scp_DM-FB}
\end{table}

\begin{table}[H]
\centering
\caption{Size-Adjusted Power, Conditional-Rolling DGP, DM-EWC}
\scalebox{0.7}{
\begin{tabular}{cc|ccccc|ccccc|ccccc}
\toprule
\toprule
&       & \multicolumn{5}{c|}{$h=1$}            & \multicolumn{5}{c|}{$h=3$}            & \multicolumn{5}{c}{$h=12$} \\
\cmidrule{3-17}          &       & \multicolumn{5}{c|}{$P$}              & \multicolumn{5}{c|}{$P$}              & \multicolumn{5}{c}{$P$} \\
$R$   & $\widetilde{R}$ & 25    & 75    & 125   & 175   & 1000  & 25    & 75    & 125   & 175   & 1000  & 25    & 75    & 125   & 175   & 1000 \\
\midrule
25 & 25& \textbf{0.050}& \textbf{0.050}& \textbf{0.050}& \textbf{0.050}& \textbf{0.050}& \textbf{0.050}& \textbf{0.050}& \textbf{0.050}& \textbf{0.050}& \textbf{0.050}& \textbf{0.050}& \textbf{0.050}& \textbf{0.050}& \textbf{0.050}& \textbf{0.050}\\
25 & 75& 0.059& 0.051& 0.035& 0.033& 0.020& 0.080& 0.066& 0.055& 0.046& 0.018& 0.068& 0.071& 0.062& 0.066& 0.044\\
25 & 125& 0.065& 0.074& 0.055& 0.042& 0.022& 0.104& 0.097& 0.073& 0.064& 0.020& 0.077& 0.106& 0.091& 0.092& 0.058\\
25 & 175& 0.076& 0.084& 0.062& 0.051& 0.023& 0.104& 0.126& 0.097& 0.072& 0.025& 0.085& 0.141& 0.140& 0.113& 0.074\\
\midrule
75 & 25& 0.057& 0.097& 0.152& 0.188& 0.577& 0.051& 0.088& 0.127& 0.167& 0.544& 0.044& 0.060& 0.084& 0.126& 0.425\\
\bottomrule
\bottomrule
\end{tabular}
}
\label{tab:CR_scp_DM-EWC}
\end{table}

\begin{table}[H]
\centering
\caption{Size-Adjusted Power, Conditional-Rolling DGP, DM-IM2}
\scalebox{0.7}{
\begin{tabular}{cc|ccccc|ccccc|ccccc}
\toprule
\toprule
&       & \multicolumn{5}{c|}{$h=1$}            & \multicolumn{5}{c|}{$h=3$}            & \multicolumn{5}{c}{$h=12$} \\
\cmidrule{3-17}          &       & \multicolumn{5}{c|}{$P$}              & \multicolumn{5}{c|}{$P$}              & \multicolumn{5}{c}{$P$} \\
$R$   & $\widetilde{R}$ & 25    & 75    & 125   & 175   & 1000  & 25    & 75    & 125   & 175   & 1000  & 25    & 75    & 125   & 175   & 1000 \\
\midrule
25 & 25& \textbf{0.050}& \textbf{0.050}& \textbf{0.050}& \textbf{0.050}& \textbf{0.050}& \textbf{0.050}& \textbf{0.050}& \textbf{0.050}& \textbf{0.050}& \textbf{0.050}& \textbf{0.050}& \textbf{0.050}& \textbf{0.050}& \textbf{0.050}& \textbf{0.050}\\
25 & 75& 0.059& 0.045& 0.040& 0.049& 0.050& 0.066& 0.048& 0.042& 0.044& 0.044& 0.052& 0.053& 0.055& 0.056& 0.059\\
25 & 125& 0.059& 0.041& 0.033& 0.038& 0.052& 0.075& 0.049& 0.034& 0.041& 0.051& 0.055& 0.056& 0.049& 0.052& 0.064\\
25 & 175& 0.062& 0.044& 0.042& 0.037& 0.053& 0.079& 0.052& 0.039& 0.035& 0.049& 0.057& 0.058& 0.050& 0.047& 0.061\\
\midrule
75 & 25& 0.060& 0.080& 0.091& 0.074& 0.131& 0.051& 0.068& 0.061& 0.072& 0.132& 0.045& 0.060& 0.062& 0.059& 0.096\\
\bottomrule
\bottomrule
\end{tabular}
}
\label{tab:CR_scp_DM-IM2}
\end{table}

\begin{table}[H]
\centering
\caption{Size-Adjusted Power, Conditional-Rolling DGP, DM-IM5}
\scalebox{0.7}{
\begin{tabular}{cc|ccccc|ccccc|ccccc}
\toprule
\toprule
&       & \multicolumn{5}{c|}{$h=1$}            & \multicolumn{5}{c|}{$h=3$}            & \multicolumn{5}{c}{$h=12$} \\
\cmidrule{3-17}          &       & \multicolumn{5}{c|}{$P$}              & \multicolumn{5}{c|}{$P$}              & \multicolumn{5}{c}{$P$} \\
$R$   & $\widetilde{R}$ & 25    & 75    & 125   & 175   & 1000  & 25    & 75    & 125   & 175   & 1000  & 25    & 75    & 125   & 175   & 1000 \\
\midrule
25 & 25& \textbf{0.050}& \textbf{0.050}& \textbf{0.050}& \textbf{0.050}& \textbf{0.050}& \textbf{0.050}& \textbf{0.050}& \textbf{0.050}& \textbf{0.050}& \textbf{0.050}& \textbf{0.050}& \textbf{0.050}& \textbf{0.050}& \textbf{0.050}& \textbf{0.050}\\
25 & 75& 0.071& 0.036& 0.025& 0.022& 0.055& 0.090& 0.052& 0.038& 0.029& 0.049& 0.069& 0.064& 0.056& 0.053& 0.057\\
25 & 125& 0.078& 0.041& 0.028& 0.018& 0.049& 0.116& 0.073& 0.032& 0.026& 0.048& 0.076& 0.087& 0.058& 0.050& 0.060\\
25 & 175& 0.093& 0.053& 0.029& 0.014& 0.043& 0.123& 0.091& 0.037& 0.024& 0.039& 0.089& 0.112& 0.089& 0.058& 0.065\\
\midrule
75 & 25& 0.062& 0.105& 0.147& 0.183& 0.332& 0.050& 0.092& 0.113& 0.154& 0.316& 0.042& 0.061& 0.077& 0.096& 0.220\\
\bottomrule
\bottomrule
\end{tabular}
}
\label{tab:CR_scp_DM-IM5}
\end{table}

\begin{table}[H]
\centering
\caption{Size-Adjusted Power, Conditional-Rolling DGP, DM-IM10}
\scalebox{0.7}{
\begin{tabular}{cc|ccccc|ccccc|ccccc}
\toprule
\toprule
&       & \multicolumn{5}{c|}{$h=1$}            & \multicolumn{5}{c|}{$h=3$}            & \multicolumn{5}{c}{$h=12$} \\
\cmidrule{3-17}          &       & \multicolumn{5}{c|}{$P$}              & \multicolumn{5}{c|}{$P$}              & \multicolumn{5}{c}{$P$} \\
$R$   & $\widetilde{R}$ & 25    & 75    & 125   & 175   & 1000  & 25    & 75    & 125   & 175   & 1000  & 25    & 75    & 125   & 175   & 1000 \\
\midrule
25 & 25& \textbf{0.050}& \textbf{0.050}& \textbf{0.050}& \textbf{0.050}& \textbf{0.050}& \textbf{0.050}& \textbf{0.050}& \textbf{0.050}& \textbf{0.050}& \textbf{0.050}& \textbf{0.050}& \textbf{0.050}& \textbf{0.050}& \textbf{0.050}& \textbf{0.050}\\
25 & 75& 0.080& 0.062& 0.039& 0.027& 0.044& 0.097& 0.073& 0.047& 0.035& 0.047& 0.082& 0.075& 0.059& 0.060& 0.051\\
25 & 125& 0.096& 0.081& 0.052& 0.033& 0.029& 0.131& 0.108& 0.064& 0.044& 0.033& 0.090& 0.106& 0.082& 0.076& 0.044\\
25 & 175& 0.116& 0.096& 0.061& 0.034& 0.018& 0.139& 0.133& 0.081& 0.048& 0.021& 0.107& 0.140& 0.123& 0.089& 0.039\\
\midrule
75 & 25& 0.056& 0.106& 0.153& 0.180& 0.425& 0.047& 0.086& 0.127& 0.155& 0.386& 0.042& 0.065& 0.097& 0.123& 0.298\\
\bottomrule
\bottomrule
\end{tabular}
}
\label{tab:CR_scp_DM-IM10}
\end{table}

\end{document}